# Magnetic Properties of Star-Forming Dense Cores

*short title:*  Dense Core Magnetic Properties

Philip C. Myers
Center for Astrophysics | Harvard and Smithsonian (CfA), Cambridge, MA 02138, USA
pmyers@cfa.harvard.edu

Shantanu Basu
University of Western Ontario (UWO), London, Ontario N6A 3K7, Canada

**Abstract**

Magnetic and energetic properties are presented for 17 dense cores within a few hundred pc of the Sun. Their plane-of-sky field strengths $B_{\text{pos}}$ are estimated from the dispersion of polarization directions, following Davis, Chandrasekhar and Fermi (DCF). Their ratio of mass to magnetic critical mass is $0.5 \lesssim M/M_B \lesssim 3$, indicating nearly critical field strengths. The field strength $B_{\text{pos}}$ is correlated with column density $N$ as $B_{\text{pos}} \propto N^p$, where $p = 1.05 \pm 0.08$, and with density $n$ as $B_{\text{pos}} \propto n^q$, where $q = 0.66 \pm 0.05$. These magnetic properties are consistent with those derived from Zeeman studies (Crutcher et al. 2010), with less scatter. Relations between virial mass $M_V$, magnetic critical mass $M_B$, and Alfvén amplitude $\sigma_B/B$ match the observed range of $M/M_B$ for cores observed to be nearly virial, $0.5 \lesssim M/M_V \lesssim 2$, with moderate Alfvén amplitudes, $0.1 \lesssim \sigma_B/B \lesssim 0.4$. The $B - N$ and $B - n$ correlations in the DCF and Zeeman samples can be explained when such bound, Alfvénic, and nearly-critical cores have central concentration and spheroidal shape. For these properties, $B \propto N$ because $M/M_B$ is nearly constant compared to the range of $N$, and $B \propto n^{2/3}$ because $M^{1/3}$ is nearly constant compared to the range of $n^{2/3}$. The observed core fields which follow $B \propto n^{2/3}$ need not be much weaker than gravity, in contrast to core fields which follow $B \propto n^{2/3}$ due to spherical contraction at constant mass (Mestel 1966). Instead, the nearly critical values of $M/M_B$ suggest that the observed core fields are nearly as strong as possible, among values which allow gravitational contraction.

# 1. Introduction

The last two decades have seen a significant expansion of observations and modeling of the effects of magnetic fields in star formation. In one interpretation, magnetic fields are strong in low-density regions (Pattle et al. 2017, Tritsis et al. 2018). They prevent gravitational contraction on large scales due to nearly flux-frozen magnetic fields, until it is initiated by ambipolar diffusion (Mouschovias & Ciolek 1999; Shu et al. 1999). This process can be aided by turbulence (Nakamura & Li 2005; Basu et al. 2009). It leads to collapsing cores, disks, outflows, and star formation. In another picture, supercritical magnetic fields influence and inhibit star formation by forming filaments, guiding flows, and by driving protostellar outflows (McKee 1999; Krumholz & Federrath 2020), but they are not a controlling influence. Measurements of Zeeman splitting of spectral lines, and maps of dust polarization have suggested that field strengths are weaker than the critical value by a factor of 2-3 (Crutcher et al. 2010, hereafter C10; Crutcher 2012, hereafter C12). If so, fields in dense star-forming regions are not strong enough to prevent gravitational collapse, but they are among the strongest which allow collapse. A more complete picture depends on accurate estimates of field strength, and on corroboration by independent methods.

To better understand the role of magnetic fields in star-forming regions, several observational techniques have been developed in addition to Zeeman splitting (Ward-Thompson et al. 2020). Improvements in near-infrared and submm instruments for polarimetry have enabled studies of star-forming clouds and cores based on polarization directions and intensity, using single-dish telescopes (Pattle & Fissel 2020, hereafter PF20) and interferometric arrays (Hull & Zhang 2020). The most widely used technique assumes that statistical fluctuations in the direction of polarization are due to turbulent excitation of Alfvén waves (Davis 1951; Chandrasekhar & Fermi 1954; DCF). The plane-of-the-sky field strength is then

$$B_{\text{pos}} = Q(4\pi\rho)^{1/2}\sigma_{NT}\sigma_\theta^{-1} \qquad (1)$$

for correction coefficient $Q$, mean density $\rho$, nonthermal velocity dispersion $\sigma_{NT}$, and dispersion of polarization directions $\sigma_\theta$. The dispersion $\sigma_\theta$ can be expressed as the ratio of field fluctuation amplitude to mean field value, $\sigma_\theta = \sigma_B/B$ (Ostriker et al. 2001, hereafter OSG01).

DCF observations are easier to carry out than Zeeman observations, but the resulting field strength estimates are less certain, due to radiative transfer effects, turbulent field structure, and the statistical nature of the method. Simulations of turbulent magnetic fields have guided estimates of the correction coefficient $Q \approx 0.5$ (Heitsch et al. 2001, Padoan et al. 2001, OSG01, Kudoh & Basu 2003). Modifications of the DCF formula have been proposed to account for large-scale fields and line-of-sight variations, via structure function analysis (Hildebrand et al. 2009, Houde et al. 2009).

It is desirable to compare DCF and Zeeman estimates of field strength in similar regions, with similar resolution and sensitivity to field strength. Such comparisons can reveal where the methods agree and disagree, and can refine the procedures of DCF analysis. DCF estimates can further investigate the distribution of the mass-to-flux ratio in star-forming regions, to test consistency with the Zeeman finding $M/M_B = 2 - 3$. They can investigate the dependence of the field strength on column density and density, and they can clarify whether fields are "strong" or "weak" compared to self-gravity in star-forming gas.

Progress toward these goals has been limited. Eight sources having both OH Zeeman and dust polarization observations were analyzed to estimate the typical ratio $B_{pos}/B_{los} = 4.7 \pm 2.8$, which is greater than the value $\pi/2$ expected from the average over orientations (Pillai et al. 2016). It is expected that some departures may arise from systematic differences in line-of-sight field structure and in analysis methods (Poidevin et al. 2013). The high-mass cold clump G35.20w in W48 was observed at similar size scales in the CN Zeeman effect and in dust polarization, yielding closer estimates of $B_{pos} = 740 \pm 250 \ \mu G$ and $B_{los} = 690 \pm 420 \ \mu G$, whose ratio is $B_{pos}/B_{los} = 1.1 \pm 0.8$ (Pillai et al. 2016).

More than 60 single-dish dust-emission polarization observations have been reported for sources which differ from those in the OH and CN sample of C10. Their polarization data have been analyzed with either the "classic" or "structure function" DCF techniques. In the log $B$ - log $n$ plane, these DCF data for $B_{pos}$ partly overlap the Zeeman data for $B_{los}$, with an order-of-magnitude scatter (PF20 Figure 2). A linear fit to the data in PF20 lies a factor ~2 below the best-fit Zeeman line of C10, with a shallower slope $0.5 \pm 0.1$ and with a correlation coefficient $R = 0.5$. This moderate level of agreement between Zeeman and DCF results calls for more specific source selection and more uniform application of DCF methods.

This paper presents a study of 17 carefully selected low-mass cores analyzed with the classic DCF technique. The goals are to determine whether DCF and Zeeman magnetic properties are consistent, to understand the distribution of $M/M_B$, and to understand the $B-N$ and $B-n$ correlations noted above.

In this paper, Section (2) describes core selection and basic core properties. Section (3) shows that the cores are generally gravitationally bound and have a supercritical mass-to-flux ratio. The cores show correlations between field strength $B$, column density $N$, and density $n$ similar to those found in Zeeman studies. Section (4) ascribes the narrow range of $M/M_B$ to gravitational binding and to a limited range of Alfvén amplitudes. Section (5) explains the $B-N$ and $B-n$ correlations as arising from bound, nearly critical cores with Alfvénic motions and spheroidal shapes. Section (6) shows that the cores in both samples are centrally concentrated. Section (7) summarizes and discusses the paper, and Section (8) presents the main conclusions. Appendix A shows that two power-law relations between $B$ and $n$ are equivalent. It then shows that the consequent relation between velocity dispersion and density is consistent with the DCF core data. Appendix B provides further information about the DCF core observations.

## 2. Core data
### 2.1. Selection of the core sample

The cores were selected to have relatively simple and similar structure, to minimize systematic errors in applying the DCF formula in equation (1). They generally lie within several hundred pc of the Sun. They have a single local maximum of column density, a velocity structure dominated by random motions rather than large-scale velocity gradients, and a polarization map which includes the position of the column density peak. Their polarization angle dispersions are less than $\sigma_{\theta,max} = 29°$, in accord with the condition for using $Q = 0.5$ (OSG01). They are primarily "low-mass" cores, either starless or associated with a small number of young low-mass stars, except for IC5146 cl47, which resembles a small cluster-forming "hub" (Myers 2009). In contrast, many of the C10 sources are "massive" cores associated with H II regions and young clusters.

Following these selection criteria, ten clouds with DCF estimates in the *Planck* Gould Belt survey (Planck Collaboration XXXV 2016) were excluded because their polarization angle dispersion exceeds $\sigma_{\theta,\max}$. This large dispersion may arise from composite polarization structure, which is seen in several large Gould Belt clouds extending more than 10° (Myers & Goodman 1991). This exclusion does not apply to small cores within the *Planck* clouds, such as Per B1. These cores have simpler polarization structure than their much larger host clouds.

Six DCF cores or clouds associated with high-mass clusters such as G240.3 (Qiu et al. 2014) were also excluded. Several of these regions have large velocity gradients, indicating that their nonthermal line widths are broadened by systematic flows. Their protostellar luminosities exceed $\sim 10^4\ L_\odot$, suggesting that their density and field structures are subject to significant distortion by ionization and winds. These cores are typically more distant than the low-mass cores in the sample by a factor $\gtrsim 10$, causing significantly coarser linear resolution for the same angular resolution.

The selected core properties were taken as given by the published papers, with slight adjustments for consistency. All cores are assumed to have the same mean mass per particle, $m = 2.33 m_H$, assuming one He atom for every five $H_2$ molecules (Kauffmann et al. 2008). Mean density estimates are based on the observed column density map, assuming spherical symmetry. All cores are assumed to have the same magnetic critical mass coefficient, $c_\Phi = (2\pi)^{-1}$ (Nakano & Nakamura 1978).

**2.2. Basic core data**

Table 1 lists mean densities $n$, mean column densities $N$, plane-of-sky magnetic field strengths $B_{\text{pos}}$, and mass-to-critical mass ratios $M/M_B$ of the selected cores, based on the original references, with slight adjustments as noted above. The ratio $M/M_B$ is also referred to as the mass-to-flux ratio. The cores are listed in order of increasing mean density. The range of mean density is set mainly by the range of instrumental resolution and field of view used to observe the cores, which are centrally concentrated. The lowest-density estimates come from telescopes with wide-field cameras (e.g. SIRPOL on the 1.4m IRSF telescope at the South African Astronomical Observatory, observing BHR71 by Kandori et al. 2020*a*)). The highest-density estimates come from interferometric arrays (e.g. the SMA observing IRAS 16293A by Rao et al. (2009). Additional information about the core observations is given in Table 7 in Appendix B.

**Table 1**
Core Properties

| (1) Core | (2) $\log n$ (cm$^{-3}$) | (3) $\log N$ (cm$^{-2}$) | (4) $\log B_{pos}$ ($\mu$G) | (5) $M/M_B$ | (6) $M/M_G$ | (7) $M/M_{GBP}$ | (8) Ref. |
|---|---|---|---|---|---|---|---|
| BHR 71 | 3.4 | 21.4 | 0.94 | 1.9 | 0.93 | 0.95 | 1 |
| FeSt 1-457 | 4.3 | 21.9 | 1.4 | 1.6 | 0.88 | 0.85 | 2 |
| Lup I C4 | 4.4 | 22.0 | 1.7 | 1.0 | 1.5 | 0.78 | 3 |
| B68 | 4.6 | 22.0 | 1.4 | 2.4 | 0.61 | 0.82 | 4 |
| B335 | 4.8 | 22.2 | 1.3 | 3.2 | 0.90 | 1.2 | 5 |
| Per B1 | 5.2 | 22.7 | 2.1 | 2.2 | 1.5 | 1.3 | 6 |
| L183 | 5.5 | 22.5 | 1.9 | 1.9 | 0.91 | 0.93 | 7 |
| L43 | 5.6 | 22.6 | 2.2 | 1.4 | 1.1 | 0.85 | 8 |
| IC5146 cl47 | 5.7 | 23.5 | 2.7 | 2.8 | 2.9 | 1.9 | 9 |
| L1544 | 5.7 | 22.7 | 2.2 | 1.7 | 1.2 | 0.99 | 10 |
| Oph C | 5.8 | 23.1 | 2.3 | 2.7 | 0.67 | 0.92 | 11 |
| Oph B2 | 6.4 | 23.5 | 2.8 | 2.7 | 2.0 | 1.6 | 12 |
| L1521F-IRS | 6.4 | 23.0 | 2.5 | 1.6 | 0.95 | 0.90 | 13 |
| BHR71 IRS1 | 6.9 | 23.2 | 2.8 | 1.5 | 1.6 | 1.0 | 14 |
| L1157 | 7.0 | 23.2 | 3.2 | 0.78 | 0.51 | 0.45 | 15 |
| N1333 I4A[a] | 7.2 | 24.1 | 3.7 | 1.3 | 1.4 | 0.88 | 16 |
| I 16293A[b] | 7.7 | 23.7 | 3.7 | 0.59 | 0.88 | 0.45 | 17 |
| random $\sigma$ | 0.2 | 0.2 | 0.2 | 0.3 | 0.4 | 0.4 | -- |
| sample med | 5.7 | 22.7 | 2.2 | 1.7 | 0.95 | 0.92 | -- |
| IQR | 2.0 | 1.3 | 1.3 | 1.3 | 0.64 | 0.26 | -- |

**Notes.** Properties of each cloud in Column (1) are derived from observed quantities as described in the text. Columns (2), (3), and (4) give respectively the log of the mean density $n$, the column density $N$, and the plane-of-the-sky component of the magnetic field strength $B_{pos}$. Columns (5), (6), and (7) give mass ratios whose numerator is $M$. The denominator is respectively the magnetically critical mass $M_B$, the virial mass $M_G$ of a uniform spherical cloud with zero external pressure and magnetic field, and the critical virial mass $M_{GBP}$ of a cloud with a nonzero field and the maximum external pressure. The clouds are listed in order of increasing mean density. The last three rows give the typical random uncertainty, the median value, and the inter-quartile range (*IQR*) for the quantities in columns (1)-(6).
[a]N1333 I4A = NGC1333 IRAS 4A. [b]I 16293A = IRAS 16293A

References - (1) Kandori et al. 2020a; (2) Kandori et al. 2020b; (3) Redaelli et al. 2019; (4) Kandori et al. 2020c; (5) Kandori et al. 2020d; (6) Coudé et al. 2019; (7), (8), and (10) Crutcher et al. 2004; (9) Wang et al. 2019: (11) Liu et al. 2019; (12) Soam et al. 2018; (13) Soam et al. 2019: (14) Myers et al. 2020; (15) Stephens et al. 2013; (16) Girart et al. 2006; (17) Rao et al. 2009

In Table 1, the mass-to-critical-mass ratio is taken from the original reference, or from

$$\frac{M}{M_B} = \frac{G^{1/2} m N}{c_\Phi B} \qquad (2)$$

where $G$ is the gravitational constant, $m$ is the mean particle mass, and $B$ is the magnetic field strength obtained from equation (1) with $Q = 0.5$. The coefficient $c_\Phi$ was found to be relatively independent of the flux distribution over the core (Tomisaka et al. 1988; see also Strittmatter 1966; Mouschovias & Spitzer 1976). It is always similar to the value $(2\pi)^{-1}$ for a uniform sheet, which is adopted here. Where no estimate is given of the inclination $i$ between the plane of the sky and the true magnetic field direction, the field strength $B$ is estimated from $B_{pos}$ using the average over a random distribution of inclinations, $B = (4/\pi) B_{pos}$ (Pillai et al. 2016, Heiles & Crutcher 2005).

Table 1 gives two estimates of the ratio of mass to virial mass $M/M_V$, based on data in the original references. Virial equilibrium and related properties have been proposed to play an important role in setting $M/M_B$ (Arons & Max 1975, Myers & Goodman 1988 (hereafter MG88), McKee et al. 1993 (hereafter M93), Zweibel & McKee 1995). The virial mass of a uniform sphere due to its self-gravity and kinetic energy is denoted here as $M_V = M_G \equiv c_G \sigma^2 R/G$ (Spitzer 1978), where $c_G = 5$ and $R$ is the effective spherical radius. The velocity dispersion $\sigma$ is related to its thermal and nonthermal components by $\sigma^2 = \sigma_T^2 + \sigma_{NT}^2$. The virial parameter is defined by $\alpha \equiv (M/M_G)^{-1}$. The relation $\alpha < 2$ is often used as a criterion of gravitational binding (Bertoldi & McKee 1992, hereafter BM92).

In Table 1 a second estimate of the virial mass is used, to include the effects of magnetic pressure and external kinetic pressure. This virial mass has a critical value when the external pressure has its maximum equilibrium value (Spitzer 1968, McKee 1999). The critical mass is written here as $M_{GP}$ when $B = 0$ and as $M_{GBP}$ when $B > 0$. The subscripts $GP$ and $GBP$ indicate the terms in the virial equation which accompany the kinetic energy term. $M_{GP}$ is defined by $M_{GP} \equiv c_{GP}\sigma^2 R/G$, with $c_{GP} = 2.36$. When $B > 0$, $M_{GBP}$ can be approximated as $M_{GBP} \approx M_{GP} + M_B$ (McKee 1989, see also Tomisaka et al. 1988). In relation to the masses in equation (48) of McKee (1999), $M_B = M_\Phi$, $M_{GP} = M_J$, and $M_{GBP} = M_{cr}$. Definitions of the masses in Table 1 are summarized in Table 2.

For each core property, the last three rows of Table 1 summarize the typical random uncertainty as described in Section 7.2, the sample median, and the interquartile range ($IQR$). The median and $IQR$ are more robust metrics than the mean and standard deviation, for asymmetrical distributions with outliers.

**Table 2**
Magnetic and Virial Mass Definitions

| (1) Symbol | (2) meaning | (3) defining equation |
|---|---|---|
| $M_B$ | magnetic critical mass | $M_B = c_\Phi \Phi G^{-1/2}$ |
| $M_V$ | virial mass | $M_V = M_G$, $M_V = M_{GP}$, or $M_V = M_{GBP}$ |
| $M_G$ | virial mass $(K, G)$ | $M_G = 5\sigma^2 R G^{-1}$ |
| $M_{GP}$ | critical virial mass $(K, G, P)$ | $M_{GP} = 2.34\sigma^2 R G^{-1}$ |
| $M_{GBP}$ | critical virial mass $(K, G, B, P)$ | $M_{GBP} \approx M_{GP} + M_B$ |

**Notes.** Column (1) gives the mass symbol. The subscripts on $M_G, M_{GP},$ and $M_{GBP}$ refer to the virial equation terms which accompany the kinetic energy term $K$: $G$ for self-gravity, $P$ for external pressure, and $B$ for magnetic field. Columns (2) and (3) describe each virial mass, and give its defining equation. In these equations, $c_\Phi$ is the magnetic critical mass coefficient, $\Phi$ is the magnetic flux within $R$, $G$ is the gravitational constant, $\sigma$ is the one-dimensional velocity dispersion, and $R$ is the effective spherical radius.

## 3. Core Properties

### 3.1. Magnetic and Virial Mass Ratios

The cores in Table 1 are close to virial equilibrium according to each definition used, since the medians of $M/M_G$ and $M/M_{GBP}$ are respectively 1.0 and 0.9, and since each ratio has a relatively small spread, with $IQR$ values of 0.6 and 0.3. They are gravitationally bound according to the criterion of BM92, since all of the cores satisfy $M/M_G > 0.5$, or equivalently $\alpha = (M/M_G)^{-1} < 2$. The core values of $M/M_B$ are slightly magnetically supercritical, with median 1.7 and $IQR = 1.3$. Their full range spans $0.5 \lesssim M/M_B \lesssim 3$.

These properties are illustrated in Figure 1, in plots of (*a*) $M/M_B$ v. $M/M_G$ and (*b*) $M/M_B$ v. $M/M_{GBP}$. These plots show that the cores are distinctly closer to their median virial ratio when the critical virial definition is used. They show no significant difference in virial properties between cores with and without associated protostars.

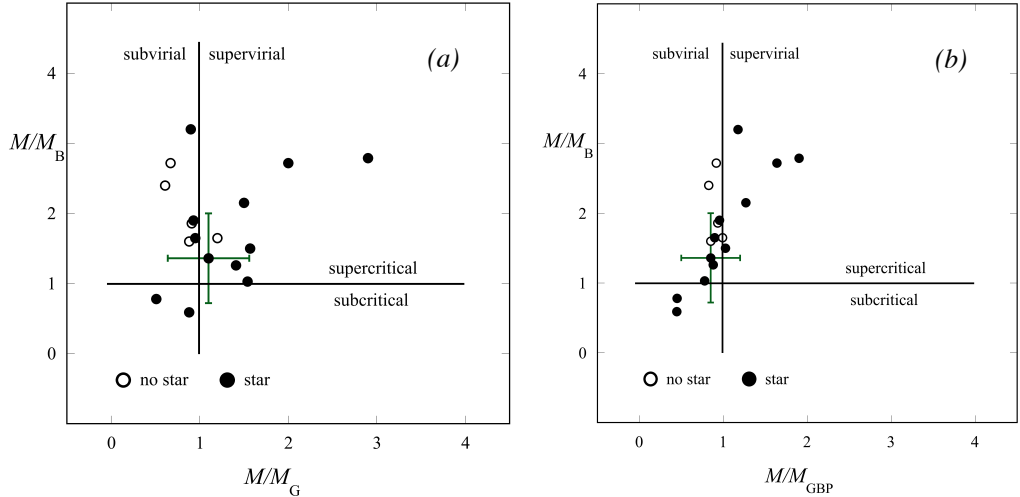

**Figure 1.** Magnetic and virial mass ratios for the cores in Table 1. Panel *(a)* shows the ratio of mass $M$ to magnetically critical mass $M_B$ as a function of the ratio of mass to virial mass. Here the virial mass $M_G$ is for a uniform sphere with negligible magnetic field and external pressure. In *(b)* the virial mass is $M_{GBP}$, the critical mass including magnetic field, at the maximum equilibrium pressure. Representative error bars are based on propagation of errors, described in Section 7.2. This figure shows that the sample cores are magnetically

supercritical and gravitationally bound, according to either virial definition, independent of whether a core has an associated protostar.

The similarity of the median values of $M/M_G$ and $M/M_{GBP}$ in Figure 1 and Table 1 can be understood from the definitions of $M_G$ and $M_{GBP}$, and because the typical core is magnetized and centrally concentrated. The central concentration at the critical external pressure causes the field-free critical mass $M_{GP}$ to be about half of the uniform virial mass $M_G$, as noted above. The internal magnetic support contributes to the overall critical mass according to $M_{GBP} \approx M_{GP} + M_B$ (McKee 1989). When the magnetic and kinetic pressure forces are comparable, $M_{GP} \approx M_B$, and consequently $M_{GBP} \approx M_G$.

Although $M/M_G$ and $M/M_{GBP}$ have similar medians, $M/M_{GBP}$ has a significantly smaller spread, with smaller $IQR$ by a factor ~2.5. This difference in spreads suggests that the typical core in Table 1 may be better described as "strongly bound" or "bound and centrally concentrated" rather than simply "bound". Further evidence for central concentration is discussed in Section 6.

In summary, the typical core in this sample may be described as strongly bound and critically virial, with $M \approx M_G \approx M_{GBP}$. Compared to the magnetic critical mass, it is supercritical by a factor ~2, so that $M/M_B \approx M/M_{GP} \approx 2$.

### 3.2. Observed relations between $B_{\mathrm{pos}}$, $N$, and $n$

The plane-of-sky field strength $B_{\mathrm{pos}}$ has strong power-law correlations with the mean core column density $N$ and with the mean core density $n$. The relation between $B_{\mathrm{pos}}$ and $N$ is shown in Figure 2, based on the data in Table 1. Figure 2 also shows the best-fit power law and the magnetically critical relation based on equation (2), assuming that $B = (4/\pi) B_{\mathrm{pos}}$.

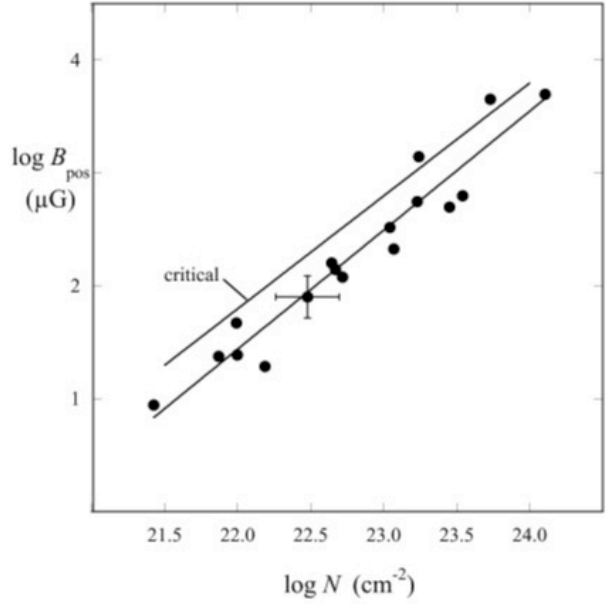

**Figure 2.** Log-log relation between plane-of-sky field strength $B_{pos}$ and mean core column density $N$ for the 17 cores in Table 1. Representative error bars are shown for the log of each variable, as discussed in Section 7.2. The equation of the best-fit line and the goodness-of-fit parameters are given in Table 3. The magnetically critical line **is** based on equation (2). This figure shows that the typical core in this sample has correlation between $B_{pos}$ and $N$, and that it is magnetically supercritical by a factor ~2.

Figure 2 shows that the typical plane-of-sky field strength is strongly correlated with the column density. The best-fit slope is consistent with unity, as would be expected for fields weaker than the magnetically critical value by a factor ~2. The fit parameters are summarized in Table 3.

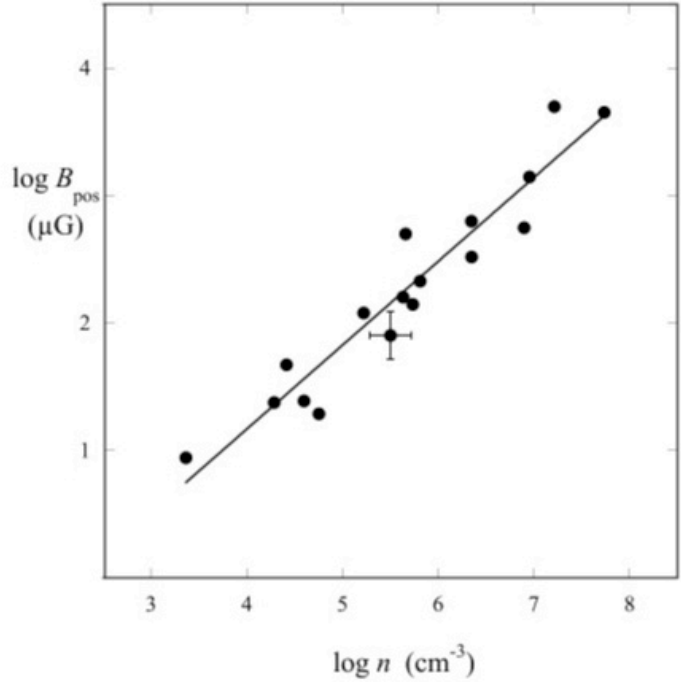

**Figure 3.** Power-law relation between plane-of-sky field strength $B_{\text{pos}}$ and mean core density $n$ for the 17 cores in Table 1. Representative error bars are shown for the log of each variable, as discussed in Section 7.2. The equation of the least-squares linear fit and the goodness-of-fit parameters are given in Table 3.

The relation between $B_{\text{pos}}$ and $n$ is shown in Figure 3. As in Figure 2, these data also show a strong power-law correlation. The equation of best linear fit and the goodness of fit parameters are given in Table 3. The slope is consistent with the value 2/3.

### 3.3. Analysis of Correlations

The power-law correlations of $B_{\text{pos}}$ with $N$ in Figure 2 and $B_{\text{pos}}$ with $n$ in Figure 3 can be understood for cores which are bound and nearly magnetically critical as shown in Figure 1, when the cores have moderate Alfvénic amplitudes and approximately spheroidal shape. Section 3.3.1 presents the basic relations used to explain these trends, and Section 3.3.2 compares the ranges of their observed variables, to account for the observed correlations.

**3.3.1. Physical relations between $B_{\rm pos}$, $N$, and $n$.** The plane-of-sky field strength is expressed in terms of the column density from the definition of the mass-to-flux ratio in equation (2) as

$$B_{\rm pos} = \frac{mG^{1/2} (\cos i) N}{c_\Phi (M/M_B)} \ . \quad (3)$$

The relation between mean core column density and density is based on a spheroidal description, because spheroids correspond to the roughly ellipsoidal intensity contours of most observed core maps. For a uniform spheroid with mass $M$, mean density $n$, and aspect ratio $Z/R$, the mean column density is

$$N = \left(\frac{16}{9\pi m}\right)^{1/3} \left(\frac{Z}{R}\right)^{2/3} M^{1/3} n^{2/3} \quad (4)$$

(Li et al. 2015, hereafter L15). Here $N$ is the mean column density along the direction of the spheroid symmetry axis, which coincides with the direction of the mean magnetic field. The length $Z$ is the spheroid radius along the symmetry axis and $R$ is the radius perpendicular to the axis. The spheroid is prolate when $Z > R$, spherical when $Z = R$, and oblate when $Z < R$.

The plane-of-sky field strength $B_{\rm pos}$ in a magnetized spheroid can be related to its mass $M$ and density $n$, in terms of the static component or the fluctuating component of the magnetic field. Eliminating $N$ from equations (3) and (4) gives $B_{\rm pos}$ in terms of the mass-to-flux ratio of the static field $M/M_B$,

$$B_{\rm pos} = c_B \left(\frac{4\pi m}{3}\right)^{2/3} G^{1/2} M^{1/3} n^{2/3} \ , \quad (5)$$

where

$$c_B \equiv 2 \cos i \, (M/M_B)^{-1} (Z/R)^{2/3} \ . \quad (6)$$

Alternately, combining the DCF expression for Alfvénic fluctuations in equation (1) with the equation for the virial mass $M_G$ in Table 2 and with the mass of a uniform spheroid, $M = (4\pi/3)mnR^2Z$, gives

$$B_{\text{pos}} = c_{\sigma_B} \left(\frac{4\pi m}{3}\right)^{2/3} G^{1/2} M^{1/3} n^{2/3} . \quad (7)$$

Here equation (7) has the same form as equation (5), but its coefficient $c_{\sigma_B}$ depends on properties $Q, \sigma_{NT}$, and $\sigma_\theta$ of the fluctuating field rather than on properties of the static field,

$$c_{\sigma_B} \equiv \left(\frac{3M_G}{5M}\right)^{1/2} \left(\frac{Z}{R}\right)^{1/6} \left(\frac{Q}{\sigma_\theta}\right) \left(\frac{\sigma_{NT}}{\sigma}\right) . \quad (8)$$

Equations (7) and (8) are consistent with equations of the form $B \propto n^{1/2} \sigma$ (Mouschovias 1991, Basu 2000, L15) due to a weak dependence of $\sigma$ on $n$, as discussed in Appendix A. Equations (5) and (7) imply that the coefficients $c_B$ and $c_{\sigma_B}$ in equations (6) and (8) are equal. Equating these coefficients relates the mass-to-flux ratio to the virial mass ratio for a spheroid with Alfvénic motions,

$$\frac{M}{M_B} = \frac{2(\cos i)\,\sigma_\theta}{Q} \left(\frac{\sigma}{\sigma_{NT}}\right) \left(\frac{5Z}{3R}\right)^{1/2} \left(\frac{M}{M_G}\right)^{1/2} . \quad (9)$$

Equation (9) can also be obtained by combining equation (1) with equation (A7) of L15.

Equation (9) confirms the typical core properties seen in Figure 1. It indicates that a bound spherical core $(M/M_G = Z/R = 1)$ with average inclination $(\cos i = \pi/4)$, transonic turbulence $(\sigma/\sigma_{NT} = \sqrt{2})$, typical polarization dispersion $(\sigma_\theta = 0.3$ radians), and standard DCF coefficient $(Q = 0.5)$ should be slightly supercritical, with $M/M_B = 1.7$. This value matches the median observed value in Table 1. The corresponding range of $M/M_B$ is discussed in Section 4.

**3.3.2. Comparison of ranges of variables.** The observed correlations between $B_{pos}$ and $N$ in Figure 2 and between $B_{pos}$ and $n$ in Figure 3 can be analyzed in terms of the relative ranges of the "primary" and "secondary" variables in equations (3) - (9). In each of equations (3), (5), and (7) $B_{pos}$ is proportional to the product of two or more variables. When one of these variables has a much greater range of values than the range of all the others, $B_{pos}$ may be correlated with that primary variable, if all of the variables are independent of each other, and if no other variables are important. Then the ranges of the secondary variables contribute to the scatter in the correlation.

The range of a variable $x$ can be quantified by the ratio of its maximum and minimum values over the sample, i.e. $r(x) \equiv x_{max}/x_{min}$. The range of a variable is equal to the range of its inverse, i.e. $r(x) = r(x^{-1})$. The comparison of ranges is then analogous to the comparison of coefficients of regressors in multivariate linear regression analysis.

Thus according to equation (3), $B_{pos}$ will appear correlated with $N$ if the range of $(\cos i)(M/M_B)^{-1}$ is much less than the range of $N$. Similarly, in equations (5) and (6), $B_{pos}$ will appear correlated with $n^{2/3}$ if the range of $(\cos i)(M/M_B)^{-1}(Z/R)^{2/3}M^{1/3}$ is much less than the range of $n^{2/3}$.

These conditions for correlation are met by the DCF sample. In equation (3) the range of $(M/M_B)^{-1}$ is 5.4, and the range of $\cos i$ can be assumed to be of order unity. Each of these ranges is much less than the range of $N$, which is 480. In equations (5) and (6), the ranges of $(M/M_B)^{-1}$ and of $\cos i$ are once again 5.4 and ~1. The range of $M^{1/3}$ is 13, and the range of $(Z/R)^{2/3}$ can also be assumed to be of order unity, as long as the core ensemble is not dominated by extremely flattened and elongated shapes. Thus each of these four ranges is much less than the range of $n^{2/3}$, which is 820. Consequently the range of $c_B M^{1/3}$ in equation (7) is much less than the range of $n^{2/3}$. Furthermore, equation (9) implies that the range of $c_{\sigma_B} M^{1/3}$ is also much less than the range of $n^{2/3}$.

In summary, the $B - N$ and $B - n$ correlations can be ascribed to core variables having distinctly different relative ranges. The $B - N$ correlation requires cores whose range of $(M/M_B)^{-1}$ is small compared to the range of $N$. This range of $(M/M_B)^{-1}$ (which is equal to the range of $M/M_B$), is associated with bound cores having modest Alfvén amplitudes, as shown in Section 4. The $B - n$ correlation requires spheroidal cores whose range of $M^{1/3}$ is small compared to the range of $n^{2/3}$. This range of $M^{1/3}$ is associated with bound cores whose density profiles are steeper than $n \propto r^{-1}$, as is shown in Section 6.

**3.4. Comparison with Zeeman results**

The statistical properties of the DCF core sample are in general agreement with those of the pioneering Zeeman studies of C10 and related papers, summarized in C12 and in Crutcher & Kemball (2020). The Zeeman cores and clouds comprise regions of high-mass and low-mass star formation, and they have very little overlap with the DCF sample. The main Zeeman findings are (1) the typical ratio of mass to magnetic critical mass is $M/M_B = 2 - 3$; and (2) the line-of-sight field strength $B_{\text{los}}$ has power-law correlations with the column density, and with the density, with exponents $\approx 1$ and $2/3$ respectively.

In this section, these Zeeman properties are compared with those based on the DCF analysis in Sections 3.1 and 3.2. Figure 4 shows side-by-side plots of *(a)* $\log B_{\text{pos}}$ *vs.* $\log N$ as in Figure 2, and *(b)* $\log B_{\text{los}}$ *vs.* $\log N$ based on the OH and CN Zeeman data in Crutcher et al. (1999), Troland & Crutcher (2008), Falgarone et al. (2008) and C10. Similarly, Figure 5 shows side-by-side plots of *(a)* the DCF relation $\log B_{\text{pos}}$ *v.* $\log n$ as in Figure 3, and *(b)* the Zeeman relation $\log B_{\text{los}}$ *v.* $\log n$ based on the foregoing references. Each plot also shows a least-squares linear fit to the data. The fit properties are summarized in Table 3.

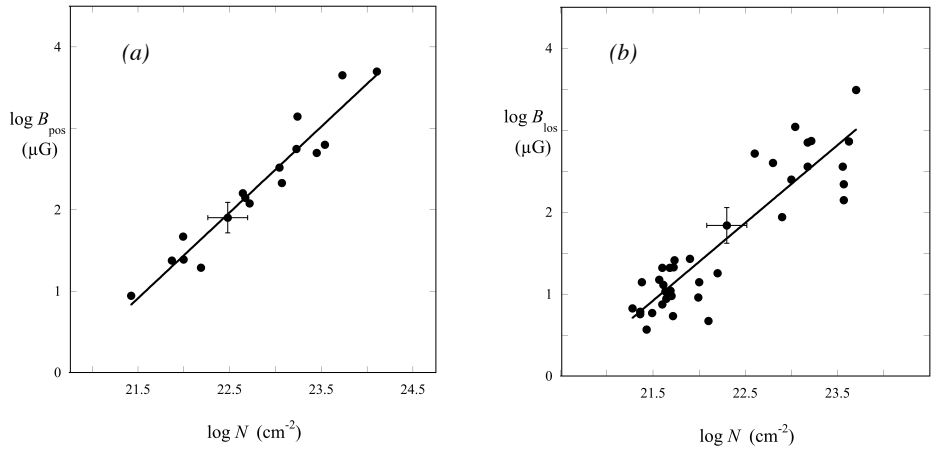

**Figure 4.** *(a)* $\log B_{\text{pos}}$ *v.* $\log N$ as in Figure 2, and *(b)* $\log B_{\text{los}}$ *v.* $\log N$ based on the OH and CN Zeeman data in Crutcher (1999), Troland & Crutcher (2008), Falgarone et al. (2008) and Crutcher et al. (2010). The Zeeman plot includes all reported values of $B_{los}$ with signal-to-noise ratio $SNR \geq 1$. The error bar on the Zeeman plot corresponds to the typical $SNR = 3$.

Table 3
Field Strength, Column Density and Density Fits

| (1) | (2) | (3) | (4) | (5) | (6) | (7) |
|---|---|---|---|---|---|---|
| Fit equation | $a$ | $\sigma_a$ | $b$ | $\sigma_b$ | $\chi^2$ | $R$ |
| $\log B_{pos} = a + b \log N$ | -21.7 | 1.7 | 1.05 | 0.08 | 0.74 | 0.96 |
| $\log B_{los} = a + b \log N$ | -19.5 | 1.7 | 0.95 | 0.07 | 4.5 | 0.91 |
| $\log B_{pos} = a + b \log n$ | -1.47 | 0.30 | 0.66 | 0.05 | 0.89 | 0.96 |
| $\log B_{los} = a + b \log n$ | -0.99 | 0.20 | 0.64 | 0.05 | 4.2 | 0.91 |

**Notes.** Column (1) gives the equations of the linear fits in Figures *4(a), 4(b), 5(a),* and *5(b).* Equations for $B_{pos}$ are for DCF data; equations for $B_{los}$ are for Zeeman data. In these equations the units of $B, N,$ and $n$ are $\mu G, cm^{-2}$, and $cm^{-3}$, respectively. Columns (2)-(7) give the best-fit parameters, their 1-$\sigma$ uncertainties, the $\chi^2$ measure of the fit quality, and the correlation coefficient $R$, according to the Levenberg-Marquardt algorithm (Levenberg 1944; Marquardt 1963).

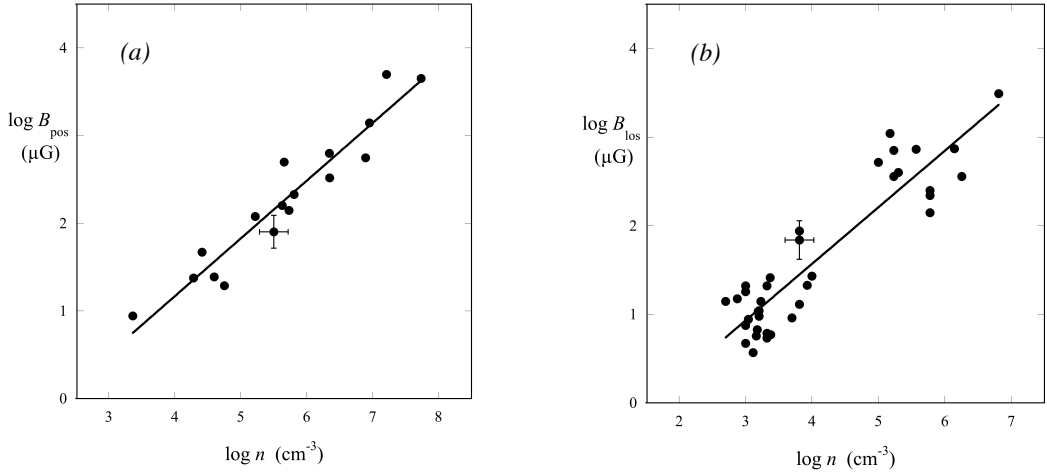

**Figure 5.** *(a)* $\log B_{\text{pos}}$ *vs.* $\log n$ as in Figure 3, and *(b)* $\log B_{\text{los}}$ *vs.* $\log n$ based on the OH and CN Zeeman data in Crutcher et al. (1999), Troland & Crutcher (2008), Falgarone et al.(2008) and Crutcher et al. (2010), as in Figure 4. The best-fit line equations are given in Table 3.

Figures 4 and 5 and Table 2 verify that the Zeeman cores and clouds have significant power-law correlations between $\log B_{\text{los}}$ and $\log N$, as is evident in C12 Figure 7, and between $\log B_{\text{los}}$ and $\log n$, as shown in C10 and in L15. Table 2 shows that the Zeeman and DCF fit equations have consistent slopes and intercepts, within $1\sigma - 2\sigma$ uncertainty. This agreement is significantly closer than between the $\log B_{\text{los}} - \log n$ fit in L15 and the $\log B_{\text{pos}} - \log n$ fit to the sample in PF20 Figure 2, described in Section 1 above. This improvement in consistency is probably due to the more careful selection of the present DCF sample, and to the use of only one method of DCF analysis. The correlation quality is discussed further in Section 7.4.

The consistency between the Zeeman and DCF fits supports the main Zeeman results of slightly supercritical fields, and approximate power-law relations $B \propto N$ and $B \propto n^{2/3}$. However, the fit uncertainties are too great to allow a useful comparison with the expected average of $B_{\text{pos}}/B_{\text{los}}$ over all orientations (Heiles & Crutcher 2005, Pillai et al. 2016).

According to Table 3, $B_{\text{los}}$ exceeds $B_{\text{pos}}$ at the same $n$ by a factor of 1.3 - 6.9, taking fit values of $a$ and $\sigma_a$ into account. If the DCF and Zeeman samples have the same trend between the total field $B$ and $n$, this range of factors corresponds to inclination angles 52 - 82 deg. However the two samples do not necessarily have identical $B - n$ trends since they have very few sources in common. A more accurate comparison would require that the same sample of objects be observed by both the DCF and Zeeman methods.

The current level of consistency does not justify revision of the standard DCF procedure or modification of the $Q$ parameter in equation (1). The mean of the best-fit values of the density exponent is slightly better-defined than by either method alone, i.e. $\bar{q} = (q_{\text{DCF}} + q_{\text{Zeeman}})/2 = 0.65 \pm 0.04$. It should be kept in mind however, that these fit results are still sensitive to sample selection, to the distribution of field directions, and to the distribution of core aspect ratios.

## 4. Supercritical values of $M/M_B$

The results in Section 2 provide the first detailed estimate of the range of values $M/M_B$ in a sample composed largely of similar, low-mass, star-forming cores. The values span the range $\mathbf{0.5} \lesssim M/M_B \lesssim 3$ with median value $M/M_B = 1.7$ and $IQR = 1.3$. For comparison the Zeeman sample used in Figures 4 and 5 can estimate $M/M_B$ by assuming $B_{\text{los}}/B$ has mean value 1/2 (Heiles & Crutcher 2005), yielding $M/M_B = 1.3$ and $IQR = 1.6$. These DCF and Zeeman values of $M/M_B$ have significant overlap, and together they strengthen the case for values of $M/M_B$ in the approximate range 1-3. This section discusses the physical origin of this relatively narrow range of supercritical values.

The supercritical values of $M/M_B$ have been discussed in terms of the virial theorem and related relationships. Observations of molecular cloud line widths, sizes and column densities suggest that cloud support against self-gravity may be due to similar contributions from magnetic and kinetic energy (MG88). Many star-forming clouds and cores are centrally concentrated, suggesting that their pressure contrast is close to that of critical virial equilibrium (Tomisaka et al. 1988). The corresponding critical virial mass can be approximated as $M_{GBP} \approx M_{GP} + M_B$ as discussed in Section 2. Then when $M \approx M_{GBP}$ and $M_{GP} \approx M_B$, $M/M_B \approx 2$ (McKee 1989). It has also been argued that $M/M_B$ may not exceed ~2 by too large a factor for large-scale cloud turbulence, since then super-Alfvénic turbulence would lead to wave dissipation in shocks and field amplification (M93; see also Boynton & Torkelson 1996 and Kudoh & Basu 2003).

It is now possible to compare the foregoing relationships among the virial mass, the mass-to-flux ratio and the Alfvén amplitudes with the observed DCF data, to show values of the physical properties which are consistent with the observed range of $M/M_B$. The DCF equation (1) is combined with the definitions of the magnetic critical mass and of the virial mass as in Section 3.3. This treatment includes thermal and nonthermal components, in cores whose nonthermal motions are subsonic, transonic, and supersonic. It is assumed that the observed polarization dispersion $\sigma_\theta$ is equal to the Alfvén amplitude ratio $\sigma_B/B$ (DCF; OSG01). The two forms of the virial mass $M_G$ and $M_{GBP}$ are considered here, as in Section 2.

Combining the field strength and magnetic critical mass equations (1) and (2) with the equations in Table 2 gives expressions for $M/M_B$ in terms of the virial mass ratios $M/M_G$ and $M/M_{GBP}$. In terms of $M/M_G$,

$$\frac{M}{M_B} = \pi \left(\frac{5M}{3M_G}\right)^{1/2} \left(\frac{\sigma_B}{B}\right)\left(\frac{\sigma}{\sigma_{NT}}\right) . \tag{10}$$

This expression is a special case of equation (9) for spherical core shape ($Z/R = 1$) with average inclination ($\cos i = \pi/4$), and standard DCF coefficient ($Q = 0.5$).

In terms of $M/M_{GBP}$,

$$\frac{M}{M_B} = \frac{2c_{GP2}}{\left(1+4c_{GP2}\left(\frac{M}{M_{GBP}}\right)^{-1}\right)^{1/2}-1} \tag{11}$$

where

$$c_{GP2} \equiv \left(\frac{\pi^2}{3}\right) c_{GP1} \left(\frac{\sigma_B}{B}\right)^2 \left(\frac{\sigma}{\sigma_{NT}}\right)^2 . \tag{12}$$

Equations (10)- (12) are identities which match the observed values of $M/M_B$ to the observed values of $\sigma_B/B = \sigma_\theta$, $\sigma/\sigma_{NT}$, and the virial mass ratios $M/M_G$ and $M/M_{GBP}$. Figure 6 uses these equations to show that the spread of values of $M/M_B$ is closely associated with gravitational binding and with the excitation and dissipation of Alfvénic fluctuations, for both subsonic and supersonic motions.

Figure 6 shows that as the velocity dispersion becomes more supersonic, the range of mass-to-flux ratios in the sample becomes independent of the velocity dispersion, and according to equation (10) it approaches the constant value $(M/M_B)_{max}/(M/M_B)_{min} = [(M/M_G)_{max}/(M/M_G)_{min}]^{1/2}(\sigma_{\theta max}/\sigma_{\theta min})$. For virial ratios close to unity as in the present sample, the range of $M/M_B$ approaches the range of $\sigma_\theta$. The range of $\sigma_\theta$ is limited, since the standard DCF method does not apply to values $\sigma_\theta \gtrsim 25°$ (OSG01), and since observed angle dispersions are rarely less than $10°$ (only three of the 17 cores in Table 7 have $\sigma_\theta < 10°$). Thus variation of the observed mass-to-flux ratio by a factor of a few should be expected for most samples of self-gravitating cores analyzed with the DCF method.

In Figure 6, the curves for $M/M_B$ as a function of $M/M_{GBP}$ (*solid lines*) bound the data more closely than the curves for $M/M_B$ as a function of (*dotted lines*). This difference suggests that the properties of the critical magnetized, centrally condensed core are more realistic for this sample of cores than those of the uniform field-free core, as discussed in Section 3.1.

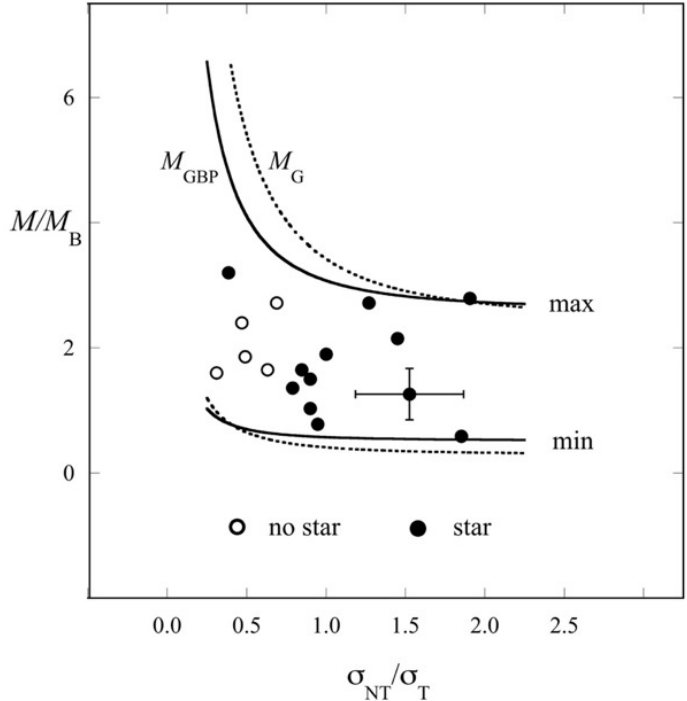

**Figure 6.** Observed values of the mass-to-critical-mass ratio for the starless cores *(open circles)* and protostellar cores *(filled circles)* in Table 1, as a function of their ratio of nonthermal and thermal velocity dispersions $\sigma_{NT}/\sigma_T$. *Curves* are predicted by equations (10)-(12), based on the DCF equation (1) and on the definitions of the magnetic critical mass $M_B$ and the virial mass, either $M_G$ *(dotted line)* or $M_{GBP}$ *(solid line)*. The upper and lower curves assume respectively the maximum and minimum observed values of $M/M_G$, $M/M_{GBP}$ and $\sigma_\theta$, using data from Tables 1 and 7. The curves closely limit the data, showing that gravitational binding and a limited range of Alfvén amplitudes are associated with the observed range of $M/M_B$.

In summary, these DCF data verify the close relationships between the static and fluctuating parts of the magnetic field and the gravitational binding of DCF cores, expected from the foregoing equations (10)-(12). These data illustrate the identities between mass-to-flux ratio, virial ratio, and polarization angle dispersion with finer detail than in earlier discussions (MG88, M93). They indicate that most DCF cores are slightly supercritical and strongly bound, sharing nearly equal contributions from their gravitational, kinetic, and magnetic energies. The range of mass-to-flux ratios is limited to a factor of a few, as expected for self-gravitating DCF samples, which have small ranges of the virial ratio, and of the dispersion in polarization angles.

## 5. Properties of the correlations $B \propto N$ and $B \propto n^{2/3}$

Section 3.3 gave a statistical explanation for the observed correlations in the DCF sample between $B$ and $N$, and between $B$ and $n^{2/3}$. This explanation is based mainly on the relative ranges of the variables $M/M_B$, $M^{1/3}$, $N$, and $n^{2/3}$ in equations (3) and (5). Section 3.4 showed that the $B - N$ and $B - n$ trends in the DCF sample resemble those in the Zeeman studies of C10. This section shows that the relative range explanation of these trends also applies to the Zeeman sample, although it has more scatter. This section also shows that the variables in the $B - N$ and $B - n$ trends are consistent with the gravitational binding properties described in Section 3.1.

All seventeen of the DCF cores were shown to be bound according to the criterion $\alpha < 2$ (BM92) in Table 1 and Figure 1. Of the 48 Zeeman cores in the OH and CN surveys, 33 also have $\alpha < 2$ (Troland & Crutcher 2008, Falgarone et al. 2008). Therefore gravitational binding is a well-defined property of the more homogeneous DCF sample, and is also prevalent in the Zeeman sample.

Table 4a presents the ranges as defined in section 3.3 of the primary variable $N$ and the secondary variable $(M/M_B)^{-1}$ for the $B - N$ correlation, in each sample. It also gives the typical ratio of each variable to its "virial counterpart" to test the relevance of gravitational binding. Here $(M/M_B)^{-1}$ has virial counterpart $M_B/M_G$ and $N$ has counterpart $M_G/(\pi m R^2)$, where $M_G = 5\sigma^2 R/G$ as in Section 2. If a primary or secondary variable is consistent with gravitational binding, the typical ratio of the variable to its virial counterpart should be close to unity.

Since $B_{pos}$ and $B_{los}$ are perpendicular components of the mean field strength $B$, each of the model equations (3), (5), (6), and (9) for $B_{pos}$ must be modified to obtain $B_{los}$ by replacing $\cos i$ with $\sin i$. This change has negligible effect on the relative range analysis, since the statistical variation in $\cos i$ over the DCF sample and in $\sin i$ over the Zeeman sample are each expected to be much smaller than the variations in the primary variables $N$ and $n^{2/3}$.

**Table 4a**
Properties of the Field Strength - Column Density Relation

| (1) | (2) | (3) | (4) | (5) |
|---|---|---|---|---|
| Sample | $r(N)$ | $\langle N/N_G \rangle$ | $r((M/M_B)^{-1})$ | $\langle M_G/M \rangle$ |
| DCF | 480 | 0.92 | 5.4 | 1.1 |
| Zeeman | 8800 | 1.6 | 35 | 1.3 |

**Notes.** This table gives observed properties of the factors in equation (3), which models the observed correlations $B_{pos} \propto N^p$ for the DCF sample and $B_{los} \propto N^p$ for the Zeeman sample. Column (1) gives the data source. Column (2) gives the observed range of the primary variable $N$, $r(N) \equiv N_{max}/N_{min}$. Column (3) gives the median ratio of $N$ to its virial value, using the definition of $M_G$ in Table 2. Columns (4) and (5) give the range and median virial

ratio for the secondary variable $(M/M_B)^{-1}$. In Column (5), the heading for the median virial ratio $\langle (M/M_B)^{-1}/(M_G/M_B)^{-1}\rangle$ is written in the simpler form $\langle M_G/M\rangle$.

Table 4a shows that the correlations between $B_{pos}$ and $N$ for the DCF sample, and between $B_{los}$ and $N$ for the Zeeman sample are each consistent with the relative ranges of the variables in equations (3), (5) and (6). The ranges of $N$ and of $(M/M_B)^{-1}$ have ratios greatly exceeding 1, since $r(N)/r((M/M_B)^{-1}) = 480/5.4 = 89$ for the DCF sample and $8800/35 = 250$ for the Zeeman sample. The typical value of each primary and secondary variable is also close to its typical virial value. The median ratio of each variable to its virial counterpart are $\langle N/N_G\rangle$ and $\langle (M/M_B)^{-1}/(M_G/M_B)^{-1}\rangle = \langle M_G/M\rangle$. Each of these ratios lies within a factor 1.1 of unity for the DCF sample, and within a factor 1.6 of unity for the Zeeman sample.

The relative ranges of $r((M/M_B)^{-1})$ and $r(N)$ in Table 4a suggest that the correlation between $B_{pos}$ and $N$ has no spurious component due to a correlation between $(M/M_B)^{-1}$ and $N$. If the relative ranges of $(M/M_B)^{-1}$ and $N$ were consistent with a power-law correlation between $(M/M_B)^{-1}$ and $N$, the relation of $B_{pos}$ to $N$ for the DCF sample would become $B_{pos} \propto N^{1+w}$ where $w = [\log r(M/M_B)]/[\log r(N)] = 0.27$. The resulting exponent 1.27 is unlikely to be consistent with the exponent $1.05 \pm 0.08$ derived from the observations. Furthermore, a linear fit between $\log((M/M_B)^{-1})$ and $\log N$ for the DCF cores has slope $0.06 \pm 0.07$ with correlation coefficient $R = 0.2$, again indicating that $(M/M_B)^{-1}$ is independent of $N$. Analysis of the Zeeman sample indicates a similar independence. A linear fit between $\log((M/M_B)^{-1})$ and $\log N$ has slope $-0.05 \pm 0.07$, with correlation coefficient $R = 0.1$.

Table 4b presents the range and virial ratio as in Table 4a, for the primary variable $n^{2/3}$ and the secondary variable $M^{1/3}$ for the $B - n$ correlations in each sample. Here the virial counterparts are $n_G^{2/3} = [3M_G/(4\pi m)]^{2/3} R^{-2}$ and $M_G^{1/3}$. The range and median virial ratio for the secondary variable $(M/M_B)^{-1}$ are given in Table 4a, so they are not repeated in Table 4b.

## Table 4b
### Properties of the Field Strength - Density Relation

| (1) Sample | (2) $r(n^{2/3})$ | (3) $\langle (n/n_G)^{2/3} \rangle$ | (4) $r(M^{1/3})$ | (5) $\langle (M/M_G)^{1/3} \rangle$ |
|---|---|---|---|---|
| DCF    | 820 | 0.95 | 13  | 0.98 |
| Zeeman | 430 | 0.58 | 9.8 | 0.92 |

**Notes.** This table gives observed properties of the factors in equations (5) and (6), which model the observed correlations $B_{pos} \propto n^q$ for the DCF sample, and $B_{los} \propto n^q$ for the Zeeman sample. Column (1) gives the data source. Column (2) gives the observed range of the primary variable $n^{2/3}$, $r(n^{2/3}) \equiv (n_{max}/n_{min})^{2/3}$. Column (3) gives the median ratio of $n^{2/3}$ to its virial value, using the definition of $M_G$ in Table 2. Columns (4) and (5) give the range and median virial ratio for the secondary variable $M^{1/3}$.

The ratios of primary to secondary ranges for the $B - n$ correlations in Table 4b greatly exceed 1 as in Table 4a, consistent with the strong correlations of $B_{pos}$ and $B_{los}$ with $n$, each approximating $B \propto n^{2/3}$. For $n^{2/3}$ and $M^{1/3}$ the ratio of ranges is $820/13 = 63$ (DCF) and $430/9.8 = 44$ (Zeeman). For $n^{2/3}$ and $(M/M_B)^{-1}$ (whose range is given in Table 4a), the ratio of ranges is $820/5.4 = 150$ (DCF) and $430/35 = 12$ (Zeeman). The median ratio of each variable to its virial counterpart lies within a factor 1.1 of unity for the DCF sample, and within a factor 1.7 of unity for the Zeeman sample. The ratio of each variable to its virial counterpart deviates from unity by nearly the same factor for the $B - n$ correlation as for the $B - N$ correlation. For both the $B - N$ and $B - n$ correlations, the Zeeman sample has significantly larger deviation from virial than the DCF sample, but neither deviation is greater than a factor of 1.7.

The correlation between $B$ and $n^{2/3}$ has no evidence of a contribution due to correlation between $M^{1/3}$ and $n^{2/3}$, according to the same analysis of relative ranges and power-law fits described above for $(M/M_B)^{-1}$ and $N$. The ranges in Table 4b and the linear fits with low correlation coefficients both indicate that $M^{1/3}$ and $n^{2/3}$ are statistically independent for both the DCF and Zeeman samples.

In summary, Tables 4a and 4b show that $(M/M_B)^{-1}$ has a smaller range than $N$, and that $(M/M_B)^{-1}$ and $M^{1/3}$ have smaller ranges than $n^{2/3}$, for the DCF and Zeeman samples. For each variable pair, the variable ranges and their correlation fits indicate that the primary and secondary variables are statistically independent. Furthermore, each factor in the $B - N$ equation (3) and in the $B - n$ equation (5) is typically close to that of its virial counterpart, within a factor of 1.1 (DCF) or 1.7 (Zeeman). Altogether the $B - N$ and $B - n$ correlations appear statistically well-defined, and the typical values of their variables appear consistent with gravitational binding.

## 6. Density structure of cores in the DCF and Zeeman samples

The foregoing results suggest that gravitational binding plays an important role in the $B - N$ and $B - n$ correlations, but they do not indicate whether these correlations are expected for all bound cores, or only for some bound cores. Section 6.1 shows that the cores in an ensemble can display a correlation of the form $B \propto n^{2/3}$ only if they are "strongly bound," or if their power-law dependence of density on spheroidal radius is steeper than $n \propto R^{-1}$. This section also shows that if these ensemble cores are bound, the $B \propto n^{2/3}$ correlation also requires that the range of velocity dispersion from core to core must be small compared to the range of radii. Finally, Section 6.2 shows that the observed DCF and Zeeman cores meet these conditions.

### 6.1. Power-law density model

Section 3.3 showed that when $B \propto M^{1/3} n^{2/3}$, a correlation of the form $B \propto n^{2/3}$ requires the range of $n^{2/3}$ to be much greater than the range of $M^{1/3}$. To apply this condition, it is assumed that an ensemble of spheroidal cores has a large range of spheroidal radii $R$, so that $r(R) \equiv R_{\max}/R_{\min} \gg 1$. It is assumed that this range of radii is not dominated by the range of aspect ratios, i.e. the range of $Z/R$ is much less than the range of $R$. It is then assumed that the mean core density $n$ within radius $R$ follows a power law $n \propto R^{-s}, s \geq 0$ from core to core. For this power law, the ratio of ranges is $r(n^{2/3})/r(M^{1/3}) = r(R^{s-1})$.

The condition for $B \propto n^{2/3}$ is then $r(R^{s-1}) \gg 1$, which is satisfied when $s > 1$. In contrast, the condition for correlation $B \propto M^{1/3}$ is $r(R^{s-1}) \ll 1$, which is satisfied when $0 \leq s < 1$. As $s$ approaches the intermediate case $s = 1$, $B$ is correlated with neither $n$ nor $M$.

It is now assumed that the cores in the ensemble have internal power-law density profiles with the same exponent from core to core. For a core of radius $R$, the internal density follows $n_{in} \propto R_{in}^{-s_{in}}$ where $0 < R_{in} \leq R$, and where the exponent $s_{in}$ is constant as $R$ varies from core to core. Then integration over $R_{in}$ to obtain the mean density within $R$ shows that $s_{in} = s$. Thus the above condition $s > 1$ for the correlation $B \propto n^{2/3}$ implies that $s_{in} > 1$ is also a condition for $B \propto n^{2/3}$. This result implies that in an ensemble which displays $B \propto n^{2/3}$ the typical core should have a "strongly-concentrated" internal density structure. In contrast, "weakly-concentrated" cores with $0 \leq s_{in} < 1$ may exhibit a correlation between $B$ and $M^{1/3}$ but not between $B$ and $n^{2/3}$, since then the range of $M^{1/3}$ would exceed that of $n^{2/3}$. As an extreme example, an ensemble of uniform density spheres ($s = 0, n \equiv n_0$) with constant $M/M_B$ would have $B \propto n_0 R \propto M^{1/3}$. Cores with $s_{in} \approx 1$ should show no correlation with either $M^{1/3}$ or $n^{2/3}$.

When gravitationally bound spheroidal cores follow $M = M_G$ as defined in Table 2, their mean density depends on velocity dispersion and radius as $n \propto \sigma^2 R^{-2}$ from core to core. Their internal density varies as $n_{in} \propto \sigma_{in}^2 R_{in}^{-2}$ within a core, where the velocity dispersion $\sigma_{in} = \sigma$ is constant within the core radius $R$, but is free to vary from core to core. To obtain $B \propto n^{2/3}$, the concentration requirement $s_{in} > 1$ can be met if the range of $\sigma$ is small compared to the range of $R$, so that $s \approx 2$. If instead $\sigma$ and $R$ have comparable ranges and if $\sigma$ varies with $R$ as $\sigma \propto R^t$, then $s$ and $t$ are related by $s = 2 - 2t$. In this case $t$ must satisfy $0 \leq t < 1/2$ to keep the core densities sufficiently concentrated, in the range $1 < s \leq 2$. For example bound cores with $\sigma \propto R^{1/2}$ would have $n \propto R^{-1}$, and thus they should display no correlation between $B$ and $n^{2/3}$ or between $B$ and $N$.

When gravitationally bound spheroidal cores are critically bound and follow $M = M_{GBP}$ instead of $M = M_G$ as assumed above, their velocity dispersion has essentially the same requirement in order to obtain $B \propto n^{2/3}$ as when $M = M_G$, provided their mass-to-flux ratio is supercritical and has small variation from core to core.

In summary, gravitationally bound cores with a large range of radii can account for the observed $B - N$ and $B - n$ correlations if their typical density profile is sufficiently concentrated, so that their gravitational binding is close to critical. If the internal density profile is a power law, the profile must be steeper than $n_{in} \propto R_{in}^{-1}$. If the velocity dispersion increases as a power of the radius as $\sigma \propto R^t$, the exponent must satisfy $t < 1/2$ so that the power-law density dependence on radius lies between $n_{in} \propto R_{in}^{-1}$ and $n_{in} \propto R_{in}^{-2}$.

## 6.2. Observed density power laws

To examine how the DCF and Zeeman samples compare to the above power-law requirements, Figure 7 presents linear fits between the log density $n$ and the log radius $R$ for each sample. Each sample shows a statistically significant power-law fit. The fit parameters are given in Table 5. The fit exponents $s$ are statistically consistent with each other, and with $s = 2$.

On the other hand, in each sample the velocity dispersion is uncorrelated with radius, so that effectively $t \approx 0$. The lack of correlation is seen in the DCF sample, where the ratio of ranges is $r(\sigma)/r(R) = 0.05$, and where a linear fit of $\log \sigma$ as a function of $\log R$ has slope -0.1 ± 0.1 and correlation coefficient 0.2. For the Zeeman sample, the ratio of ranges is 0.2, and the linear fit has slope -0.2 ± 0.1, with correlation coefficient 0.2. Therefore the DCF and Zeeman samples each have velocity dispersion exponents $t \approx 0$.

With these exponent values $s \approx 2$ and $t \approx 0$ the DCF and Zeeman cores are consistent with central concentration and strong gravitational binding. This property gives a physical basis for the statistical requirement $r(M^{1/3}) \ll r(n^{2/3})$ for the correlation $B \propto n^{2/3}$ discussed above.

These samples of cores differ from larger-scale samples of "clouds" with $R \gtrsim 0.1$ pc having $s \approx 1$ and $t \approx 1/2$ (Heyer & Dame 2015, Larson 1981).

The central concentration of cores was not an explicit criterion of selection for either the DCF or the Zeeman samples. The simplest explanation appears to be that the cores in each sample were selected to be bright enough to be detected, and that bright cores tend to be centrally concentrated. Such selection is discussed further in Section 7.4.

The $\log n$ vs. $\log R$ data for the Zeeman sample in Figure 7b has a significant linear fit consistent with $s = 2$ for all the data, but not for the subsample of CN cores with $n > 10^4$ cm$^{-3}$, and not for the subsample of OH dark clouds with $n < 10^4$ cm$^{-3}$. The whole sample has $s = 2.2 \pm 0.2$ with correlation coefficient 0.8, while the CN sample has $s = 0.7 \pm 0.4$ with coefficient 0.5, and the OH sample has $s = 0.7 \pm 0.1$ with coefficient 0.5. According to the above analysis, such weakly-concentrated samples with $s < 1$ should not display a significant correlation between $B$ and $n$.

Indeed, neither of these subsamples shows a significant $B - n$ correlation. Least-squares fitting between $\log B_{los}$ and $\log n$ gives $B_{los} \propto n^q$, $q = 0.04 \pm 0.67$ for the CN sample and $q = 0.40 \pm 0.45$ for the OH sample (L15). The plot of $\log B_{los}$ vs. $\log N$ in Figure 4b also shows correlation for the entire sample, but not within each of its subsamples. These results are consistent with the above power-law analysis for a sample with weak concentration. However, studies with larger samples are needed to determine how much this lack of correlation can be ascribed to weak concentration, and how much to small-sample statistics.

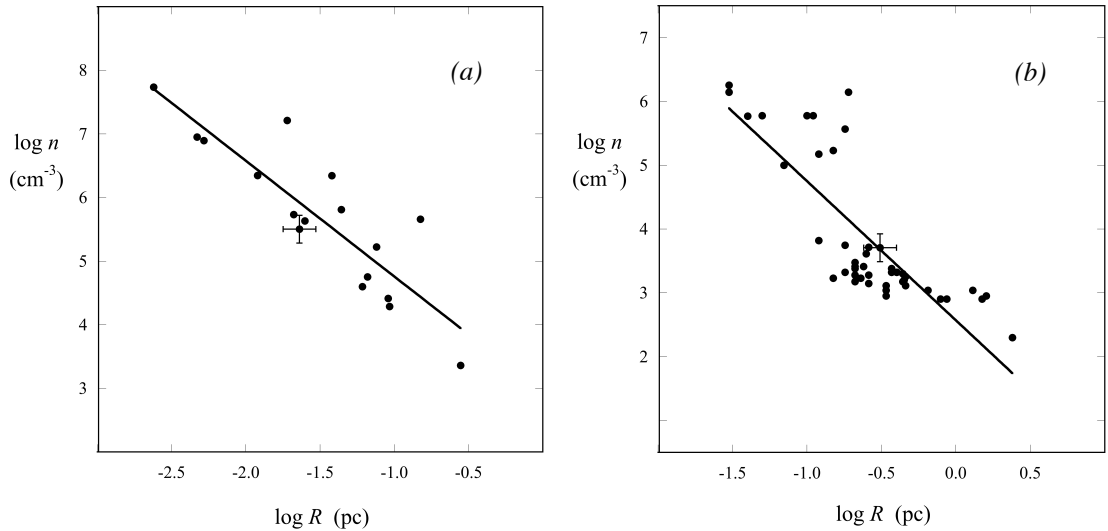

**Figure 7.** Log-log plots of mean core density $n$ versus core radius $R$, for (*a*) the DCF cores in Table 1, and (*b*) the Zeeman cores in C10. *Straight lines* indicate linear fits, whose parameters are given in Table 5. These fits to the dependence of mean density on radius from core to core are interpreted in terms of the typical density profile within a core, assuming that the cores have the same internal density profile.

**Table 5**
Density-Radius Fits of Form $\log n = a - q \log R$

| (1) | (2) | (3) | (4) | (5) | (6) | (7) |
|---|---|---|---|---|---|---|
| Sample | $a$ | $\sigma_a$ | $q$ | $\sigma_q$ | $\chi^2$ | $R$ |
| DCF | 2.9 | 0.4 | 1.8 | 0.3 | 5.3 | 0.87 |
| Zeeman | 2.6 | 0.2 | 2.2 | 0.2 | 20 | 0.82 |

**Notes.** Column (1) gives the data source. Columns (2)-(7) give the best-fit parameters, their 1-$\sigma$ uncertainties, the $\chi^2$ measure of the fit quality, and the correlation coefficient $R$, according to the Levenberg-Marquardt algorithm (Levenberg 1944; Marquardt 1963).

## 7. Discussion

This section discusses the results of the paper, with a summary in Section 7.1, limitations and uncertainties in Section 7.2, and implications and related work in Section 7.3.

### 7.1. Summary

This study was carried out to better understand the origin of three magnetic properties of star-forming dense cores: the ratio of core mass to magnetically critical mass $M/M_B$ (also known as the mass-to-flux ratio), the relation of mean field strength $B$ to mean column density $N$, and the relation of $B$ to mean density $n$. The method is based on polarization observations of dense cores within a few hundred pc, and on estimates of their magnetic properties based on the models of Davis (1951) and of Chandrasekhar and Fermi (1954) (DCF). This study is timely because there are now enough published DCF estimates to allow a statistically useful selection of well-resolved cores in nearby regions of star formation.

DCF observations of cores within a few hundred pc of the Sun were selected to obtain estimates of the mass-to-flux ratio $M/M_B$ and of the virial ratios $M/M_G$ and $M/M_{GBP}$, which respectively exclude and include the effects of internal fields and external pressure. This compilation indicates that most of the DCF cores have mass-to-flux ratio close to 2 (slightly supercritical) and virial mass ratios close to 1 (gravitationally bound). The typical spread in each mass ratio is also a factor 2-3, as shown in Table 1. These values and their narrow ranges play important roles in explaining the $B - N$ and $B - n$ correlations.

The plane-of-the-sky field strength $B_{pos}$ is strongly correlated with $N$, approximating $B_{pos} \propto N$, and $B_{pos}$ is strongly correlated with $n$, approximating $B_{pos} \propto n^{2/3}$, as shown in Figures 2 and 3. These DCF correlations for $B_{pos}$ are consistent within 1-2 $\sigma$ with Zeeman power-law correlations for $B_{los}$, shown in Figures 4 and 5. The DCF core sample is more homogeneous than the Zeeman sample, and its correlations have less scatter.

The $B \propto N$ correlations are explained statistically by expressing $B$ in terms of $(M/M_B)^{-1}$ and $N$ as in equation (3). The observed range of $(M/M_B)^{-1}$ is much smaller than the range of $N$, for both the DCF and Zeeman samples, so that $B$ appears correlated with $N$. The $B \propto n^{2/3}$ correlations are explained for spheroids by expressing $B$ in terms of $M^{1/3}$ and $n^{2/3}$ as in equations (5) and (7). The observed range of $M^{1/3}$ is much smaller than the range of $n^{2/3}$, for both samples, indicating $B \propto n^{2/3}$.

The supercritical values of the mass-to-flux ratio $M/M_B$ are analyzed by relating $M/M_B$ to the virial ratios $M/M_G$ and $M/M_{GBP}$ and to the dispersion of polarization directions $\sigma_\theta$. When these relations are applied to the maximum and minimum values of $M/M_G$, $M/M_{GBP}$, and $\sigma_\theta$, the predicted range of $M/M_B$ closely matches the observed range, as shown in Figure 6. Thus the supercritical values of $M/M_B$ can be understood in greater detail than before, verifying the importance of virial equilibrium and limited Alfvénic fluctuations, with nearly equal contributions of gravitational, kinetic, and magnetic energy (MG88, M93).

The range conditions $r((M/M_B)^{-1}) \ll r(N)$ and $r(M^{1/3}) \ll r(n^{2/3})$ which underlie the $B-N$ and $B-n$ correlations are closely associated with gravitational binding. In each sample the typical ratio of $(M/M_B)^{-1}$ to its virial value $(M_G/M_B)^{-1}$, which equals $M_G/M$, is close to unity. So is the typical ratio of $N$ to its virial value $M_G/(\pi m R^2)$, as shown in Table 4a. Similarly, in each sample the typical ratio of $M^{1/3}$ to its virial value $M_G^{1/3}$ is close to unity, as is the typical ratio of $n^{2/3}$ to its virial value $n_G^{2/3} = [3M_G/(4\pi m)]^{2/3} R^{-2}$, as shown in Table 4b.

Only strongly bound cores are expected to follow the $B-N$ and $B-n$ correlations seen in the DCF and Zeeman samples analyzed in this paper. According to a power-law density model, these correlations can arise for an ensemble of cores with a wide range of radii $R$, only if the cores are centrally concentrated, with internal density profiles steeper than $n_{in} \propto R_{in}^{-1}$. This property is demonstrated for the DCF and Zeeman cores in Figure 7. Bound cores can be strongly bound, provided their velocity dispersions have smaller ranges than their radii, as is seen in the DCF and Zeeman samples. The central concentration of these cores may be a result of selecting the cores for observation based on their brightness.

**7.2. Limitations and Uncertainties**

The main limitations on the results of this study are the heterogeneous nature of the DCF and Zeeman sources analyzed, and the variety of instruments used to estimate their column density, source size, velocity dispersion, and polarization angles, with differing sensitivity and resolution. This diversity makes it difficult to assign meaningful uncertainties to the observational quantities in the figures and tables. Instead, estimates of the typical random uncertainty in each DCF quantity are made with standard propagation of errors, by assuming that the relative uncertainty in density and column density is $\sigma_n/n = \sigma_N/N = 1/2$, the relative uncertainty in nonthermal velocity dispersion and polarization angle dispersion is $\sigma_{\sigma_{NT}}/\sigma_{NT} = \sigma_{\sigma_\theta}/\sigma_\theta = 1/4$, and the relative uncertainty in the core mass $M$, magnetic critical mass $M_B$, and virial critical mass $M_{GP}$ is $\sigma_M/M = \sigma_{M_B}/M_B = \sigma_{M_{GP}}/M_{GP} = 1/3$. In log-log plots the error bar in the log of a given quantity $x$ which has uncertainty $\sigma_x$ is computed from $\sigma_{\log x} = \sigma_x/(x \ln 10)$.

Although these error estimates are informed guesses and therefore somewhat arbitrary, they are probably not grossly wrong, since their error bars are comparable to the scatter in the highest-quality correlation plots in Figures 2 and 3. The Zeeman plot quantities are assumed to have the same random errors as their DCF counterparts. The greater scatter in the Zeeman plots than in the DCF plots is attributed to greater diversity in the selected Zeeman source properties.

The criteria of source selection can affect the conclusions drawn from statistical samples, as is shown by the significantly different correlation properties of the $\log B_{pos} - \log n$ plots between the large DCF sample in Figure 2 of PF20, described above in Section 1, and the smaller DCF sample in Figure 3. The improvement in correlation quality may be due in part to the use of stricter selection criteria in this paper. These criteria exclude candidate sources which are more likely to introduce systematic errors in evaluation of the DCF formula, as described in Section 2.1.

These results suggest that a further reduction in uncertainty may be made by undertaking new, more sensitive Zeeman and DCF polarization observations on a common set of target sources, selected as in Section 2.1. These could be observed with one Zeeman instrument and spectral line, and with one DCF instrument and polarization wavelength, instead of the many diverse instruments which are drawn on here.

### 7.3. Implications

#### 7.3.1. Weak, moderate, and strong fields.

The results in this paper add new support to the idea that fields in star-forming cores are magnetically supercritical by a factor of order 2. The DCF analysis in Sections 2 and 3 gives $0.5 \lesssim M/M_B \lesssim 3$. This result agrees with earlier Zeeman estimates (C10, C12), using an independent method of field estimation, with a different source sample, and with finer precision. Consequently these estimates are now on a firmer observational basis. This regime of field strengths is weaker than the "strong" subcritical regime, $M/M_B < 1$, where the field can prevent gravitational collapse, but it is stronger than the "weak" supercritical regime, $M/M_B \gg 1$, where the field is dynamically negligible. Thus the typical fields in the DCF and Zeeman samples are "moderate" in strength compared to gravity.

This property of $M/M_B$ appears inconsistent with a "weak-field" interpretation of the Zeeman trend $B \propto n^q$, $q \approx 2/3$ as noted by C12. The weak-field interpretation is based in part on the similarity of the exponent $q \approx 2/3$ derived from observations to the exact exponent $q = 2/3$ in the contracting core model of M66. The M66 model requires a weak field with respect to gravity to maintain its initially spherical shape as it contracts. For example, a contracting magnetized Bonnor-Ebert sphere must have a mass-to-flux ratio $M/M_B > 6$ to keep its aspect ratio less than 1.05 during its first free-fall time (Kataoka et al. 2012). This value of $M/M_B > 6$ is significantly greater than any of the values of $M/M_B$ derived from observations in Table 1.

If the $B \propto n^q$, $q \approx 2/3$ trend is instead ascribed to the scaling relations in Sections 3-6, the observed scaling can be understood in terms of an ensemble of objects (not necessarily an evolutionary sequence) that have moderately strong magnetic fields and modest variation of mass-to-flux ratios and virial parameters.

This moderate-field interpretation has support from numerical MHD simulations. Here we review some relevant studies. Many simulations have found that gravitationally-contracting dense molecular cloud regions develop a systematic scaling of the form $B \propto n^q$. Using three-dimensional (3D) nonideal MHD simulations, Kudoh et al. (2007) found that high density regions undergoing gravitational collapse in a nonturbulent sheet tend toward the $B - n$ relation with $q \approx 1/2$ and mildly supercritical values, regardless of whether the initial conditions of the cloud were subcritical or supercritical. Both the evolutionary track of the highest density cell as well as an ensemble of high-density cells at the final output time followed this trend. Hennebelle et al. (2008) simulated a collision of two streams of warm atomic gas in a 3D ideal MHD calculation and found that the $B - n$ relation was relatively flat (showing a modest trend of increase) below a density $\sim 10^3 \text{cm}^{-3}$ but increased rapidly toward $q \approx 1/2$ for higher densities, where the gas is supercritical. However the spatial resolution of $\sim 0.05$ pc cannot resolve the $B - n$ relation on the smaller scales typical of DCF cores.

Mocz et al. (2017; hereafter M17) performed very high-resolution 3D ideal MHD simulations with a range of scales from a box size of about 5 pc to a minimum cell size of about 4 au. They started their models with a range of initial normalized mass-to-flux ratio ranging from 0.8 to 80, applied driven Fourier-space turbulence that ranged from sub-Alfvénic to highly super-Alfvénic as the initial mass-to-flux ratio increased. Gravity was turned on after the driven turbulence reached a steady state and the simulation ended in less than 0.2 initial free-fall times in the moderate to strong magnetic field models due to the collapse of the highest density cells. At the final snapshot a plot of the $B - n$ relation from cell values revealed a transition to a relation of the form $B \propto n^q$ above a density of about $10^4$ cm$^{-3}$. The index $q$ is notably less than 2/3 for densities less than about $10^6$ cm$^{-3}$, corresponding to length scales $r \simeq 10^3 - 10^4$ au. Above a density of about $5 \times 10^6$ cm$^{-3}$, corresponding to length scales $r < 10^3$ au, the density cells correspond to collapsing regions and the $B - n$ relation transitions to $q \approx 2/3$ in the moderate to weak initial magnetic field models. There is significant numerical reconnection in these models and the initially moderate magnetic field model with initial normalized mass-to-flux ratio 2.7 has a final value of 12.1, signifying that it is a weak field environment at this final time. Hence, the $q \approx 2/3$ in the collapsing cores can be interpreted in the manner of M66, as isotropic collapse of gas in weak field conditions. However, the collapsing cores are of higher density than the observed cores and clumps for which data was analyzed by C10, as M17 have noted.

Perhaps the most complete attempt to connect simulations of the $B - n$ relation to observables has been made by L15, who also performed 3D ideal MHD simulations with driven turbulence having thermal Mach number $M = 10$. They ran a weak-field simulation and a moderate-field simulation, with initial normalized mass-to-flux ratios of 16 and 1.6, respectively, and with initial plasma $\beta = 0.02$ and 2, respectively. The simulation is similar in concept to that of M17; gravity is turned on after a steady-state turbulence is established and then the simulation terminates less than one free-fall time $(\tau_{\rm ff})$ later. The smallest cell size is about 500 au, more than sufficient to capture bulk properties on the observed scale of clumps measured by C10. L15 identify the 100 most massive clumps in their final snapshot and calculate a mass-averaged magnetic field in each clump, in manner meant to sample the field in the same way as Zeeman measurements, including convolution with a beam size of ~ 5000 au. There is a clear power-law correlation of the estimated total magnetic field strength with the mean density, and in the moderate field model, the index $q$ estimated from the ensemble of objects has a best-fit value $0.62 \pm 0.11$ when measured at the final output time of $0.64 \tau_{\rm ff}$. These clumps also have moderately strong magnetic field strengths with mildly supercritical mass-to-flux ratio. When a time average of $q$ during the times $(0.4-0.64)\tau_{ff}$ is taken, they obtain a value $0.70\pm0.06$, consistent with a value of about 2/3. Altogether, the L15 models and analysis point to the need to explain observed $q$ values by studying ensembles of objects rather than the evolutionary track of single objects, and by incorporating observational effects to simulations when comparing with data.

Among the four simulations discussed here, the MHD turbulent simulation of L15 comes closest to matching the supercritical field strengths and power-law dependence $B \propto n^q$, $q \approx 2/3$ on the scales of the cores observed in this study.

**7.3.2. Relation to filamentary gas.** Despite the agreement between properties of observed cores and some MHD turbulent simulations, this picture is incomplete. It does not relate these core properties to the observed association of cores and larger-scale hubs with their extensive networks of filamentary gas, as discussed in Könyves et al. (2015) and André et al. (2014). It will be important for future studies to more closely relate the present results to the magnetic properties of filaments and cores as they form, as discussed e.g. in the fragmentation studies of Inutsuka & Miyama (1997), in the converging-flow formation models of Chen & Ostriker (2015) and Chen et al. (2016), and in the analytic model of Auddy et al. (2016).

**7.3.3. Relation to constant-mass core evolution models.** The interpretation of $B \propto n^q$, $q \approx 2/3$ in terms of strong binding and nearly critical field strengths supports the inference of moderate field strength estimates $M/M_B = 2-3$, and disfavors the weak-field interpretation of $B \propto n^q$, $q \approx 2/3$ associated with the M66 model. Nonetheless the M66 model remains useful for evaluating the scaling of $B$ with $n$ in spherical cores, and in oblate and prolate spheroids (Myers et al. 2020, hereafter M20). In this case the scaling is done with respect to the current mean field strength within the resolution radius of a DCF observation, following M20 equation (28), with no need to refer to a uniform initial state.

## 7.4. Quality of $B - N$ and $B - n$ correlations

The $B - N$ and $B - n$ trends in figures 2-5 have relatively good measures of correlation, with $R > 0.9$ and $\chi^2 < 5$ according to Table 3. They have significantly better quality than the $B - n$ trend in PF20 Figure 2, which has a wide range of source properties and several DCF analysis methods, yielding $R = 0.5$. They also have much better quality than either Zeeman trend, when the Zeeman sample is divided into high-and low-density regions (L15).

The DCF trends have significantly better statistics than the Zeeman trends, since the DCF trends have $R = 0.96$ and $\chi^2 < 0.5$ while the Zeeman trends have $R = 0.91$ and $\chi^2 = 4.2 - 4.5$ according to Table 3. Also, each DCF trend is statistically consistent with the original DCF sample when the DCF sample is divided into cores denser and less dense than the median density. It is therefore of particular interest to understand the basis of these higher-quality $B - N$ and $B - n$ correlations. for the DCF cores.

The simplest explanation for the quality of the $B - N$ and $B - n$ correlations among the DCF cores appears to be the nature of their selection. Their selection for observation was likely biased toward simple systems observable with good signal-to-noise ratio. Thus they tend to be bright, well-defined, local maxima of column density. These objects were further selected for inclusion in this study to have similar distance of a few hundred pc from the Sun and to have simple velocity and polarization structure.

Perhaps as a result of these criteria, the selected cores appear similar in their degree of central concentration, with typical density profile $n \propto r^{-q}$, with $q = 1.8 \pm 0.3$ as given in Table 5. The selected cores are therefore strongly gravitationally bound, since their velocity dispersions are uncorrelated with radius, as noted in Section 6.2, and since their virial ratios $M/M_G$ and $M/M_{GBP}$ are each close to unity, as given in Table 1. This strong binding and the small range of observed polarization dispersions ensure that the mass to flux ratio $M/M_B$ also varies by only a factor of a few, as discussed in Sections 3 and 4.

The observational resolutions summarized in Table 7 span a factor of 14. As these resolutions are applied to centrally concentrated cores, the resulting mean densities, column densities, and field strengths span several orders of magnitude. In contrast to these extensive ranges of $n, N,$ and $B,$ the typical variation in the secondary parameters $M/M_G$, $M/M_B$, $\sigma_{NT}$, and $\sigma_\theta$ by a factor of a few is relatively small. Therefore the correlations among $n, N,$ and $B$ can have relatively high quality because they have relatively small contribution from secondary variables.

Nonetheless, it may appear surprising that a selection of cores observed in a wide variety of star-forming clouds can yield similar, well-defined $B-n$ and $B-N$ trends having less scatter than in earlier Zeeman results. While the $B-N$ trend can appear in simulations as a result of near-flux-freezing once a mildly supercritical value of $M/M_B$ is reached, the $B-n$ correlation with $\kappa \approx 2/3$ arises in cases of very specific initial conditions in some studies. For example they require highly turbulent gas with Mach number $M \gtrsim 10$ and plasma $\beta \geq 1$ in OSG01, $M \approx 10$ and Alfvén Mach number $M_A \approx 1$ and $\beta \approx 0.02$ in L15, and $M \approx 10$ and $\beta \approx 0.4$ in Chen et al. (2016). A connection between the observational $B-n$ correlation and the formation properties requires more detailed observations, as well as simulations that study a wide range of initial conditions and make predictions for ensembles of cores. A question which would be useful to address is the degree to which strongly self-gravitating cores retain "memory" of their initial environmental properties.

## 8. Conclusions

The main conclusions of this paper are:

1. Polarization observations of 17 nearby dense cores are compiled to estimate magnetic properties of cores with the DCF method.

2. The cores are on average mildly supercritical, with mass-to-flux ratio $M/M_B \simeq$ 1-3. They are strongly gravitationally bound, with ratio of mass to critical virial mass $M/M_{GBP} \simeq$ 0.5 - 2. The typical core has $M/M_{GBP} = 1$, virial parameter $\alpha = 1$, and $M/M_B = 2$. The typical Alfvén Mach number is $M_A = 0.8$, indicating comparable turbulent and magnetic energy. The prevalence of strong binding in the sample may be a consequence of selecting relatively bright cores for observation.

3. The plane-of-sky field strengths are correlated with column density $N$ as $B_{\text{pos}} \propto N^p$, where $p = 1.05 \pm 0.08$, and with density $n$ as $B_{\text{pos}} \propto n^q$, where $q = 0.66 \pm 0.05$. These properties match those derived from Zeeman studies (C10), with less scatter.

4. The range of values of $M/M_B$ in the DCF cores is consistent with strong binding and dissipation of large-amplitude Alfvén fluctuations. The observed range of $M/M_B$ is consistent with the observed ranges of critical virial mass ratio $0.5 \lesssim M/M_{GBP} \lesssim 2$ and polarization angle dispersion $5° \leq \sigma_\theta \leq 20°$, which corresponds to Alfvén fluctuation amplitudes $0.1 \leq \sigma_B/B \leq 0.4$.

5. The correlations between $B$ and $N$, and between $B$ and $n$ can be explained for a sample of strongly bound, nearly magnetically critical cores with spheroidal shape and a large range of radii. $B$ is related to $N$ and $n$ through the identities $B \propto (M/M_B)^{-1}N$ and $N \propto M^{1/3}n^{2/3}$. For nearly critical cores the range of $(M/M_B)^{-1}$ is less than the range of $N$, so $B$ correlates with $N$. For strongly bound cores, the range of $M^{1/3}$ is less than the range of $n^{2/3}$, so $B$ correlates with $n^{2/3}$.

6. This statistical explanation of $B \propto n^{2/3}$ applies to both the DCF and Zeeman samples. In contrast to the relation $B \propto n^{2/3}$ in the constant-mass contraction model of Mestel (1966), this explanation does not require fields to be very weak compared to gravity in order to allow the initial core shape to remain spherical. In each explanation the same exponent 2/3 occurs because it is the ratio of space dimensions for flux and mass.

7. The trend $B \propto n^{2/3}$ can arise for any sample of bound and nearly critical cores with a wide range of radii and column density, a small range of velocity dispersion, and density power laws which are steeper than $n \propto r^{-1}$. These conditions are satisfied by the DCF and Zeeman cores, with typical density power laws close to $n \propto r^{-2}$.

8. The relations $B \propto n^{2/3}$ and $B \propto n^{1/2}\sigma$ imply a scaling between velocity dispersion and density $\sigma \propto n^{1/6}$ as noted by L15 for Zeeman data. The relations are consistent, since $B \propto n^{1/2}\sigma$ can be derived from $B \propto n^{2/3}$ for bound cores having small variation in virial parameter and mass-to-flux ratio. The scaling $\sigma \propto n^{1/6}$ is verified for the DCF cores, whose best-fit power law matches the expected relation within one-sigma uncertainty, in coefficient and in exponent.

**Acknowledgements**

The authors thank Ian Stephens, Tyler Bourke, Sarah Sadavoy, Tom Dame, Juan Soler, and Henrik Beuther for helpful comments. P. Myers thanks Terry Marshall for support and Janice Sexton for editorial assistance. The authors thank the organizers of the 2020 Heidelberg-Harvard Workshop on Star Formation, where this work was presented in preliminary form. S. Basu is supported by a Discovery Grant from the Natural Sciences and Engineering Research Council of Canada. The authors thank the referee for prompt and constructive comments, which improved the paper.

**Appendix A**

This appendix shows that an equation of the form $B \propto n^{1/2}\sigma$ can be derived from the relation $B \propto n^{2/3}$ discussed throughout the text, for cores with small variation in the virial ratio $\alpha$ and in the mass-to-flux ratio. The scaling relation $\sigma \propto n^{1/6}$ implied by these two power laws is derived analytically for both coefficient and exponent. The DCF data are shown to follow this relation within one-sigma in both coefficient and exponent. The Zeeman data were shown to follow a similar scaling by L15.

Equations (7) and (8) express the original equation (1) of the DCF method in a form that is comparable to the form of $B_{\text{pos}}$ obtained for a spheroid in equations (5) and (6). Here we show that we can also go in the opposite direction and express the spheroidal estimate in a form similar to the DCF estimate.

Defining $\alpha \equiv M_G/M$, where $M_G \equiv 5\sigma^2 R/G$ (BM92), we obtain from equations (5) and (6) that

$$B_{\text{pos}} = 5^{1/3} \alpha^{-1/3} c_B \left(\frac{4\pi m}{3}\right)^{2/3} G^{1/6} \left(\sigma^2 R\right)^{1/3} n^{2/3}. \tag{A1}$$

Note that in the above, we could have stayed more precisely with the spheroidal (as opposed to spherical) assumption by letting $M_G = \left(5\sigma^2/G\right) R (Z/R)^{1/3}$, which would introduce an extra factor of $(Z/R)^{1/6}$ in equation (A1) that could be absorbed into the expression for $c_B$. We later use the result that $(Z/R)$ does not vary much from core to core, so this would not make a substantial difference.

Next, we use that for a spheroid $M = (4/3)\pi R^2 Z \rho = (4/3)\pi R^3 (Z/R) \rho$ and if the magnetic field is along the z-axis then $\Phi = B\pi R^2$, where $B$ and $\rho$ are mean values. Using the definition $\mu \equiv (M/M_B) = (G^{1/2}/c_\Phi)(M/\Phi)$, the mass and flux expressions can be combined to yield

$$B = \frac{4}{3}\mu^{-1} \frac{G^{1/2}}{c_\Phi}\left(\frac{Z}{R}\right)\rho R. \tag{A2}$$

Using $B_{pos} = B\cos i$ and rearranging we find that

$$R = \frac{3}{4}\frac{c_\Phi}{G^{1/2}}\left(\frac{Z}{R}\right)^{-1}\frac{\mu}{\cos i}\frac{B_{pos}}{\rho}. \tag{A3}$$

Inserting equation (A3) into (A1) we find

$$B_{pos} = \left(\frac{4}{3}\pi m\right)^{1/2} c_{B2}\sigma n^{1/2}, \tag{A4}$$

where

$$c_{B2} = (40\pi)^{1/2} \cos i\, \alpha^{-1/2} \mu^{-1} c_\Phi^{1/2} \left(\frac{Z}{R}\right)^{1/2}. \tag{A5}$$

Equation (A4) shows that the scaling $B \propto \sigma n^{1/2}$, which was shown to be a good fit to Zeeman data by Basu (2000), is an equivalent form to equation (5), as long as the virial parameter $\alpha$ along with other parameters in equation (A5) do not vary much from core to core. More precisely, equation (A4) provides a good fit if the composite dimensionless value $c_{B2}$ does not vary much from core to core. In turn, equations (A4) and (A5) are consistent with the DCF equation (1) by combining them with equation (9).

The relevance of the expression of equation (A4) in interpreting the DCF and Zeeman data can be seen in that a fit that is effectively $B \propto n^{2/3}$ implies a hidden scaling $\sigma \propto n^{1/6}$. A similar $\sigma - n$ relation was found in the Zeeman measurements of dense star-forming regions by L15 using data compiled by C12. This increase of velocity dispersion with increasing density can be understood in the Zeeman data by noting that the sample is quite heterogeneous and the high-density objects include many high mass clouds that have enhanced the velocity dispersion. Hence the Zeeman data represents an ensemble of clouds and should not be considered an evolutionary sequence.

The scaling relation $\sigma \propto n^{1/6}$ inferred above can also be derived explicitly, so that both the coefficient and the exponent can be compared with observations. Combining the DCF equation (1) with the spheroid equations (5) and (6) gives

$$\sigma_{NT} = \frac{2(4\pi m)^{1/6} G^{1/2}}{3^{2/3} Q} \cos i \left(\frac{Z}{R}\right)^{2/3} \left(\frac{M}{M_B}\right)^{-1} \sigma_\theta M^{1/3} n^{1/6} \qquad (A6)$$

where all quantities are defined in Sections 2 and 3. Then the observables $\sigma_{NT}$, $M/M_B$, $\sigma_\theta$, and $M^{1/3}$ can be gathered so that their combination $S$ depends only on $\cos i \left(\frac{Z}{R}\right)^{2/3} n^{1/6}$, i.e.

$$S \equiv \frac{\sigma_{NT}(M/M_B)}{\sigma_\theta M^{1/3}} = c_S \cos i \, (Z/R)^{2/3} n^{1/6}, \qquad (A7)$$

where $c_S = 2.08 \times 10^{-3}$ km s$^{-1}$deg$^{-1}(M_\odot)^{-1/3}$(cm$^{-3})^{-1/6}$. The quantity $\cos i \left(\frac{Z}{R}\right)^{2/3}$ can be considered to be a parameter which applies to the entire sample. Then the observables $S$ and $n$ can be plotted for the cores in the sample, for comparison with the model in equation (A7). This comparison is shown in log-log form in Figure 8.

Figure 8 shows $\log S$ vs. $\log n$ for the 17 DCF cores in Table 1. The lines indicate a least-squares linear fit and the model equation (A7) for spherical cores ($Z/R = 1$) having the statistical average inclination ($\cos i = \pi/4$). The linear fit parameters are given in Table 6.

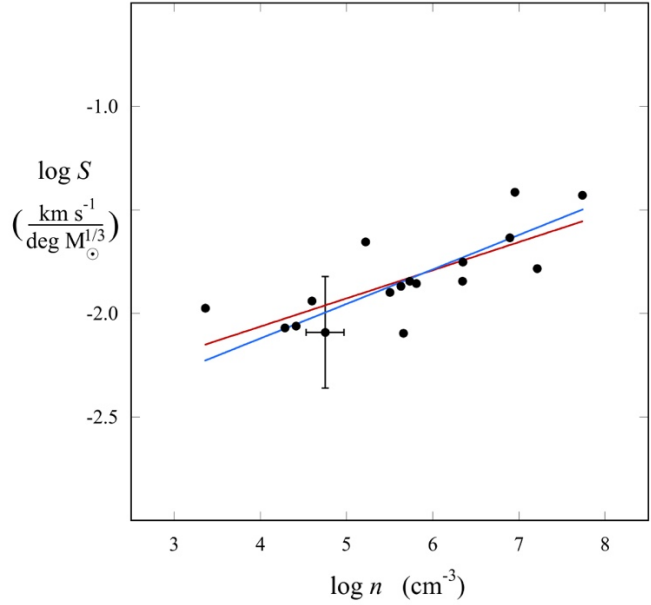

**Figure 8.** Log-log variation of nonthermal velocity dispersion function $S$ with mean density $n$ for DCF cores. The function $S$ is the combination of the nonthermal velocity dispersion and three other observables defined in equation (A7). *Filled circles* indicate the observed data for the cores in Table 1. The *red line* is the least-squares linear fit to the data, with fit parameters given in Table 6. The *blue line* is the model equation (A7), assuming that the observed cores have spherical shape and the statistical average field inclination $\cos i = \pi/4$.

Table 6
Fit and Model Parameters of Form $\log S = a + b \log n$

| (1) | (2) | (3) | (4) | (5) | (6) | (7) |
|---|---|---|---|---|---|---|
| Source | $a$ | $\sigma_a$ | $b$ | $\sigma_b$ | $\chi^2$ | $R$ |
| DCF 17 | -2.6 | 0.2 | 0.14 | 0.03 | 0.30 | 0.76 |
| DCF 16 | -2.8 | 0.2 | 0.16 | 0.03 | 0.25 | 0.79 |
| Model | -2.8 | --- | 0.17 | ---- | ---- | ---- |

**Notes.** The model function $S$ is the combination of the nonthermal velocity dispersion $\sigma_{NT}$ and three other observables, defined in equation (A7). Column (1) gives the parameter origin. DCF 17 indicates fits to all 17 cores in Table 1. DCF 16 indicates removal of the candidate outlier BHR71, whose density has the greatest deviation from the other sample members (see text). Columns (2)-(7) give the best-fit parameters, their 1-$\sigma$ uncertainties, the $\chi^2$ measure of the fit quality, and the correlation coefficient $R$, according to the Levenberg-Marquardt algorithm (Levenberg 1944; Marquardt 1963). The model parameters are due to evaluation of the log of equation (A7) for $Z/R = 1$ and $\cos i = \pi/4$.

Figure 8 and Table 6 show that the DCF cores have a linear trend in a log-log-plot, whose intercept and slope are each consistent with the model equation (A7), within 1-sigma uncertainty. The slope is therefore close to the value 1/6 expected for consistency between the relations $B \propto n^{1/2} \sigma_{NT}$ and $B \propto n^{2/3}$. This conclusion does not change if one excludes the lowest-density point, which has the greatest deviation in $\log n$ from its nearest neighbor and the greatest deviation from the mean. In that case the quality of the correlation and the agreement with the model each improve slightly, as is evident from the values in Table 6.

## Appendix B

This appendix supplements the information in Table 1 with Table 7, which gives more detailed properties of the 17 cores and their observations.

Table 7
Observed Core Properties

| (1) Core | (2) $\delta R$ (pc) | (3) $R$ (pc) | (4) $M$ ($M_\odot$) | (5) $\Delta v$ (km s$^{-1}$) | (6) $\sigma_{NT}$ (km s$^{-1}$) | (7) mol | (8) $\lambda_{pol}$ ($\mu$m) | (9) $\sigma_\theta$ (deg) | (10) PS | (11) ref |
|---|---|---|---|---|---|---|---|---|---|---|
| BHR 71 | 0.021 | 0.28 | 12 | 0.04 | 0.20 | NH$_3$ | 1.6 | 15-25 | 2 | 1,18 |
| FeSt 1-457 | 0.014 | 0.093 | 3.6 | 0.12 | 0.057 | N$_2$H$^+$ | 1.6 | 6.9 | 0 | 2,19 |
| Lup I C4 | 0.014 | 0.091 | 4.7 | 0.11 | 0.11 | N$_2$H$^+$ | 214 | 15 | 1 | 3,20 |
| B68 | 0.013 | 0.061 | 2.1 | 0.14 | 0.091 | N$_2$H$^+$ | 1.6 | 15 | 0 | 4,21 |
| B335 | 0.011 | 0.066 | 3.7 | 0.19 | 0.13 | H$_2$CO | 1.6 | 21 | 1 | 5,22 |
| Per B1 | 0.010 | 0.076 | 16 | 0.07 | 0.29 | NH$_3$ | 850 | 11 | 1 | 6 |
| L183 | 0.011 | 0.023 | 1.0 | 0.06 | 0.093 | N$_2$H$^+$ | 850 | 14 | 0 | 7,23 |
| L43 | 0.013 | 0.025 | 1.7 | 0.06 | 0.14 | N$_2$H$^+$ | 850 | 12 | 0 | 8,23 |
| IC5146 cl47 | 0.077 | 0.15 | 85 | 0.03 | 0.36 | C$^{18}$O | 850 | 17 | 4 | 9,24,25 |
| L1544 | 0.014 | 0.021 | 1.3 | 0.06 | 0.12 | N$_2$H$^+$ | 850 | 13 | 0 | 10,23 |
| Oph C | 0.013 | 0.044 | 12 | 0.06 | 0.13 | N$_2$H$^+$ | 850 | 11 | 0 | 11,26 |
| Oph B2 | 0.013 | 0.038 | 42 | 0.06 | 0.24 | N$_2$H$^+$ | 850 | 15 | 4 | 12,26 |
| L1521F-IRS | 0.014 | 0.012 | 0.84 | 0.06 | 0.16 | N$_2$H$^+$ | 850 | 15 | 1 | 13,27 |
| BHR71 IRS1 | 0.0016 | 0.022 | 2.0 | 0.08 | 0.17 | C$^{18}$O | 1300 | 17 | 1 | 14,18,28 |
| L1157 | 0.0025 | 0.0047 | 0.19 | 0.098 | 0.18 | N$_2$H$^+$ | 1300 | 6 | 1 | 15,29 |
| N1333 I4A[a] | 0.0018 | 0.0050 | 1.2 | 0.13 | 0.2 | H$_2$CO | 870 | 5 | 2 | 16,30 |
| I16293A[b] | 0.0015 | 0.0024 | 0.33 | 0.7 | 0.35 | H$^{13}$CO$^+$ | 880 | 10 | 1 | 17 |

**Notes.** This table gives observational properties of the DCF sample not already given in Table 1. For each core in column (1), column (11) refers to the primary article, spectroscopic observations and associated protostars. Column (2) is the beam radius at the source distance. Column (3) is the source radius based on a spherical model. Column (4) is the mass within $R$. Column (5) is the FWHM velocity resolution of the molecular line observation which yields the nonthermal velocity dispersion (columns 6, 7). Column (8) is the wavelength of the continuum polarization observation. Column (9) is the dispersion of polarization angles used to estimate the magnetic field strength by the DCF method. Column (10) gives the estimated number of associated protostars (either Class 0 or Class I). [a]N1333 I4A = NGC1333 IRAS 4A. [b]I16293A = IRAS 16293A

References - (1) Kandori et al. 2020a; (2) Kandori et al. 2020b; (3) Redaelli et al. 2019; (4) Kandori et al. 2020c; (5) Kandori et al. 2020d; (6) Coudé et al. 2019; (7), (8), and (10) Crutcher et al. 2004; (9)Wang et al. 2019: (11) Liu et al. 2020; (12) Soam et al. 2018; (13) Soam et al. 2020: (14) Myers et al. 2020; (15) Stephens et al. 2013; (16) Girart et al. 2006; (17) Rao et al. 2009; (18) Tobin et al. 2019; (19) Tatematsu et al. 2004; (20) Benedettini et al. 2012; (21) Lada et al. 2003; (22) Zhou et al. 1990; (23) Caselli et al. 2002; (24) Buckle et al. 2009; (25) Dunham et al. 2015; (26) André et al. 2007; (27) Lee et al. (2001); (28) Hull et al. 2020; (29) Chiang et al. 2010; (30) Di Francesco et al. 2001